\documentclass[11pt,a4paper]{article}
\pdfoutput=1
\usepackage{jheppub}
\usepackage{bm}
\usepackage{amsmath,amsfonts,amssymb,mathrsfs,graphicx}
\usepackage[squaren]{SIunits}
\usepackage{verbatim}
\usepackage{hyperref}
\usepackage{slashed}
\usepackage[utf8]{inputenc}
\usepackage{float}
\usepackage{bbold}
\usepackage{pifont}
\usepackage{url}
\usepackage[dvipsnames]{xcolor}
\usepackage{subfigure}
\usepackage[labelfont=bf, format=hang, justification=centerlast, singlelinecheck=false, font=small]{caption}
\usepackage{afterpage}
\usepackage{cleveref}
\usepackage{array}
\usepackage{rotating}
\usepackage{changes}
\normalsize

\Crefname{figure}{Fig.}{Figs.}

\title{Characterizing simplified models for heavy Higgs decays to supersymmetric particles} 

\author[a]{Suchita Kulkarni}
\author[a]{Lukas Lechner}

\emailAdd{suchita.kulkarni@oeaw.ac.at}
\emailAdd{lukas.k.lechner@gmail.com}

\affiliation[a]{Institut f\"ur Hochenergiephysik,  
\"Osterreichische Akademie der Wissenschaften, \\ Nikolsdorfer Gasse 18, 1050 Wien, Austria}

\abstract{
The search for heavy Higgs bosons is an important step to probe the parameter space of the Minimal Supersymmetric Standard Model. In this work, we classify all possible decay modes of the supersymmetric heavy Higgs boson using the \texttt{SModelS} framework. We work within the phenomenological MSSM and use the ATLAS pMSSM study as our viable parameter space. We find that for a bino-like and higgsino-like LSP, a significant region of the parameter space results in mono-$X$ ($X$ = $h$, $W$, $Z$) final states. For wino-like LSPs, we demonstrate the existence of displaced vertex signatures with a large signal cross section. Finally, we argue that by covering the mono-$X$ final states, a large part of heavy Higgs decays can be tested at the LHC.}


\begin{document}

\maketitle

\section{Introduction}\label{sec_intro}
\noindent
Among Beyond the Standard Model (BSM) searches at the Large Hadron Collider (LHC), heavy Higgs searches form an important component. While the discovery of a Standard Model (SM) Higgs boson~\cite{Aad:2012tfa, Chatrchyan:2012xdj} has brought a new era to the searches for scalar particles, there is no final word on whether the observed Higgs boson is unique or whether it is a part of an extended sector. 

Extended Higgs sectors appear in a variety of BSM scenarios~\cite{Branco:2011iw, LopezHonorez:2006gr, Drees:873465, Gunion:1989we,Djouadi:2005gj}. The primary production modes for a heavy neutral Higgs within such extended sectors are gluon fusion ($ggH$) and bottom- ($bbH$) and top-quark annihilation ($ttH$)\footnote{It is also possible to produce Higgs via vector boson fusion and Higgs-strahlung processes, however they are generally suppressed.}. At the LHC, searches for extended Higgs sectors are primarily carried out in SM final states, as key decay modes are demonstrated to be $\gamma\gamma$, $\tau \bar{\tau}$, $b\bar{b}$ and $VV$, where $V$ = $W$, $Z$, $h$. However, heavy Higgs decays to BSM particles are also possible. If the branching ratio of a heavy Higgs to BSM particles is large, the searches for SM final states may no longer be effective.  

Among the theoretical scenarios featuring extended Higgs sector, supersymmetry (SUSY) is a well motivated candidate~\cite{Drees:873465, Martin:1997ns}. While searches for the direct production of SUSY particles are extensively carried out at the LHC, it is also possible that the supersymmetric heavy neutral Higgs decays to SUSY particles. This opens another possibility to search for SUSY. 

It has been previously shown that within the R-parity conserving Minimal Supersymmetric Standard Model (MSSM), heavy Higgs bosons can in general have a large branching ratio to SUSY particles~\cite{Barman:2016kgt, Arbey:2013jla, Belanger:2015vwa, Djouadi:1996mj}, correspondingly resulting in large signal cross sections with SM and missing energy (MET) final states at the LHC~\cite{Bisset:2007mk, Arhrib:2011rp, Belanger:2000tg, Bisset:2000ud, Medina:2017bke, Barman:2016jov, Bisset:2007mi, Gunion:1988yc, Li:2013nma, Ananthanarayan:2015fwa, Djouadi:2015jea, Zhang:2002fu, Ibrahim:2008rq}. Despite these early encouraging studies, a complete classification of supersymmetric Higgs to SUSY final states and associated signatures at colliders remains to be done. 

In this paper, we aim at systematically classifying MSSM neutral heavy Higgs decays to SUSY final states in a realistic parameter space of the phenomenological MSSM (pMSSM). Our study differs from previous attempts in several ways. First and foremost, we allow for squarks and sleptons to be light, in addition to the electroweak sector focused on so far. An initial discussion in this direction can be found in~\cite{Barman:2016jov}. Secondly, we take into account the parameter space currently allowed by direct and indirect experimental searches for BSM. Finally, we make a systematic and complete survey of possible MSSM Higgs decay modes as well as discussing their implications at the LHC. As a byproduct of this study, we also demonstrate a new way of using the \texttt{SModelS}~\cite{Ambrogi:2017neo,Kraml:2013mwa,SModelS:wiki} framework for classifying BSM decays of resonances. \texttt{SModelS} is an automated tool that decomposes an input BSM scenario obeying a $\mathbb{Z}_2$ symmetry into simplified model topologies.

This paper is organized as follows: In section~\ref{sec:MSSMhiggs}, we review the MSSM Higgs sector and discuss the dependence of heavy Higgs branching ratios to SUSY particles on the underlying parameters. We detail our analysis setup in section~\ref{sec:anlysis}, followed by a discussion on the results in section~\ref{sec:results}. We conclude in section~\ref{sec:conclusion}, with an outlook.

\section{The MSSM Higgs Sector}
\label{sec:MSSMhiggs}
Supersymmetric Models contain two Higgs-doublets ($H_u$, $H_d$) to break the electroweak symmetry and therefore predict the existence of two neutral CP-even ($h$, $H$), one neutral CP-odd ($A$) and a pair of charged scalar bosons ($H^\pm$). In addition, a neutral and a pair of charged Goldstone bosons ($G^0$, $G^\pm$) are retrieved from the mixing of the real ($\phi$) and complex ($\chi$) components of the Higgs doublets using the (not independent) mixing angles $\alpha$ and $\beta$~\cite{Gunion:1989we, Djouadi:2005gj, Drees:873465}.
\begin{align}
H_d &= \begin{pmatrix}H_d^0\\H_d^-\end{pmatrix} = \begin{pmatrix}\left(\nu_d + \phi_d^0 + i\chi_d^0\right)/\sqrt{2}\\\phi_d^-\end{pmatrix},\qquad \left\langle H_d\right\rangle = \frac{\nu_d}{\sqrt{2}}\\
H_u &= \begin{pmatrix}H_u^+\\H_u^0\end{pmatrix} = \begin{pmatrix}\phi_u^+\\\left(\nu_u + \phi_u^0 + i\chi_u^0\right)/\sqrt{2}\end{pmatrix},\qquad \left\langle H_u\right\rangle = \frac{\nu_u}{\sqrt{2}}\\\nonumber\\
\begin{pmatrix}H\\ h\end{pmatrix} = \begin{pmatrix} c_\alpha & s_\alpha\\ -s_\alpha & c_\alpha\end{pmatrix} &\begin{pmatrix}\phi_d^0\\ \phi_u^0\end{pmatrix},\quad
\begin{pmatrix}G^0\\ A\end{pmatrix} = \begin{pmatrix} c_\beta & s_\beta\\ -s_\beta & c_\beta\end{pmatrix} \begin{pmatrix}\chi_d^0\\ \chi_u^0\end{pmatrix},\quad
\begin{pmatrix}G^\pm\\ H^\pm\end{pmatrix} = \begin{pmatrix} c_\beta & s_\beta\\ -s_\beta & c_\beta\end{pmatrix} \begin{pmatrix}\phi_d^\pm\\ \phi_u^\pm\end{pmatrix}\nonumber\\\nonumber\\
\cos^2\left(\beta - \alpha\right) &= \frac{m_h^2\left(m_Z^2 - m_h^2\right)}{m_A^2\left(m_H^2 - m_h^2\right)},\qquad 0 \le \beta \le \frac{\pi}{2},\qquad -\frac{\pi}{2} \le \alpha \le 0\label{eq_mixingangles}
\end{align}
Here, $s_\alpha \equiv \sin\alpha$ and $c_\alpha \equiv \cos\alpha$. At tree-level, the Higgs masses are determined by two independent parameters: the ratio of the two vacuum expectation values (vevs), $\tan\beta = v_u/v_d$, with $v_d^2 + v_u^2 = v_\text{SM}^2$, and the mass of the pseudo-scalar, $m_A$~\cite{Gunion:1989we, Djouadi:2005gj, Drees:873465}. Since the mass of the light Higgs boson cannot exceed the $Z$ boson mass for leading order calculations, radiative corrections play a crucial role in the determination of the MSSM Higgs masses. These radiative corrections arise mainly due to $3^\text{rd}$ generation (s)particles leading to a dependence on the chosen parameters for the squark and slepton masses. In the decoupling limit $m_A \gg m_Z$, $h$ is SM-like and the masses of the heavy Higgses are (almost) degenerate. Within this work, we will focus only on the phenomenology of heavy neutral MSSM Higgs boson ($H$).

\subsection*{Heavy Higgs Boson Couplings to Weakinos}
The couplings to weakinos depend on gaugino-higgsino admixture of neutralinos and charginos. Therefore, the couplings for left- and right-handed weakinos to the CP-even neutral heavy Higgs bosons
\begin{align}
g^L_{\tilde{\chi}^\pm_i\tilde{\chi}^\mp_jH} = g^R_{\tilde{\chi}^\pm_j\tilde{\chi}^\mp_iH} &= \frac{1}{\sqrt{2}s_W}\left(\cos\alpha \, \mathcal{V}_{j1} \,\mathcal{U}_{i2} + \sin\alpha \, \mathcal{V}_{j2} \, \mathcal{U}_{i1}\right)\\
g^{L,R}_{\tilde{\chi}^0_i\tilde{\chi}^0_jH} &= \frac{1}{2s_W}\left(\mathcal{Z}_{j2} - t_W \, \mathcal{Z}_{j1}\right)\left(\cos\alpha \, \mathcal{Z}_{i4} - \sin\alpha \, \mathcal{Z}_{i3}\right) + i \leftrightarrow j
\end{align}
\noindent
where $s_W$ and $t_W$ represent the sine and tangens of the weak mixing angle, are strongly dependent on the neutralino (chargino) mixing matrices $\mathcal{Z}$ ($\mathcal{U}$, $\mathcal{V}$) and therefore on the pMSSM parameters $M_1$, $M_2$, $\mu$ and $\tan\beta$~\cite{Gunion:1989we, Drees:873465}. Furthermore, due to the appearance of the Higgs doublet mixing angle $\alpha$, the couplings also depend on the Higgs boson masses, thus the parameter $m_A$. For the decoupling regime ($m_A \gg m_Z$), the mixing angle~$\alpha$ is small and terms proportional to $\sin\alpha$ become negligible.

\subsection*{Heavy Higgs Boson Couplings to Sfermions}\label{sec_sfermioncoupling}
The couplings of the heavy neutral Higgs bosons to sfermions can be large for the third generation, as they include terms proportional to their SM partner mass squared~$m^2_f$ and terms proportional to their trilinear couplings~$A_f$. Furthermore, couplings to the bottom squark are enhanced for large values of $\tan\beta$ and can exceed those to top squarks. The Lagrangian for the interaction\footnote{Here, no quartic interaction terms are considered.} of heavy Higgs bosons to squarks can be written as
\begin{align}
\mathcal{L}_{H\tilde{q}\tilde{q}} &= \sum_{\tilde{q}} C_{H\tilde{q}\tilde{q}} H\,\tilde{q}^*\,\tilde{q},\text{ with}\nonumber\\
C_{H\tilde{q}\tilde{q}} &= \begin{pmatrix}
\left(I_q^{3L} - Q_q s_W^2\right)m_Z^2\cos(\beta + \alpha) + m_q^2 r^q_1 & \frac{1}{2}m_q\left(A_q r_1^q + \mu r_2^q\right)\\
\frac{1}{2}m_q\left(A_q r_1^q + \mu r_2^q\right)	& Q_q s_W^2m_Z^2\cos(\beta + \alpha) + m_q^2 r^q_1
\end{pmatrix},\label{eq_squarkmixing}
\end{align}
where $C_{H\tilde{q}\tilde{q}}$ contains the couplings to squarks, $r_1^u = \sin\alpha/\sin\beta$, $r_2^u = -\cos\alpha/\sin\beta$, $r_1^d = -\cos\alpha/\cos\beta$ and $r_2^d = -\sin\alpha/\cos\beta$, $s_W$ represents the sine of the weak mixing angle, $I_q^{3L}$ is the weak isospin, $Q_q$ is the electric charge, $m_q$ is the mass of the quark and finally $A_q$ is the trilinear coupling. For the decoupling regime, where $m_H\gg m_Z$, the Higgs mixing angle~$\alpha$ is negligible for moderate to large $\tan\beta$ and thus, only $r_2^u$ and $r_1^d$ are non-vanishing. Furthermore, the trilinear couplings of the $1^\text{st}$ and $2^\text{nd}$ generation squarks are set to zero in pMSSM. Due to the light mass of the SM light quarks, the coupling of $1^\text{st}$ and $2^\text{nd}$ generation squarks can be approximately reduced to the diagonal terms in the matrix given above, leading to a strong dependence on $\cos\beta$, and thus the pMSSM parameter $\tan\beta$. Qualitative features of Higgs couplings to charged sleptons are similar to Higgs couplings to down-type squarks and the exact couplings will not be described here.

Additionally, the mixing angles $\alpha$ and $\beta$, and therefore the mass of the Higgs bosons (see eq.~\ref{eq_mixingangles}) influence the couplings to up-type and down-type fermions, as well as couplings to vector bosons. The coupling of the heavy Higgses ($H$, $A$) to fermions strongly depend on $\tan\beta$ leading to crucial effects in heavy Higgs production mechanisms~\cite{Gunion:1989we, Djouadi:2005gj, Drees:873465}. Due to this, at large $\tan\beta$ the bottom-quark annihilation ($bbH$) production mechanism dominates over gluon fusion ($ggH$). 

An important point to remember from this discussion is that while the heavy Higgs production cross section increases with $\tan\beta$, the branching ratio to weakinos decreases and the branching ratio to down-type squarks and charged sleptons increases. However, Higgs decay to $b\bar{b}$ is also enhanced and typically it wins over decays to sfermions for large $\tan\beta$. We will comment about this further in the results section. 

\section{Analysis Setup}
\label{sec:anlysis}
Even though a full 19-dimensional parameter scan of the pMSSM is computationally expensive, it is a necessity to systematically probe the full parameter space. This study uses the pre-collected and categorized pMSSM dataset from a multi-dimensional flat-prior scan, stored in the SLHA file format~\cite{Aad:2015baa}. It assumes soft SUSY breaking, R-parity conservation and no additional sources of CP and flavor violation at the electroweak scale. It further demands the Lightest Supersymmetric Particle (LSP) to be a stable dark matter candidate, the lightest neutralino. The SM input values and the soft parameter ranges for the scan are given in Ref.~\cite{Berger:2008cq}.

The pre-collected data sample is already tested against LEP constraints, current limits from the flavor sector, the muon $g_\mu-2$ and obeys the upper bound on the Dark Matter relic density. It also requires the SM Higgs mass to be between 124 to 128 GeV. The points passing all these constraints are checked against 22 distinct ATLAS Run 1 SUSY searches. Further details about the applied constraints are given in Ref.~\cite{Aad:2015baa}. For underlying theory studies see~\cite{Berger:2008cq, CahillRowley:2012cb, CahillRowley:2012kx, Cahill-Rowley:2014twa}. The availability of the data sample which has been thoroughly tested against existing LHC limits is a primary reason to choose the ATLAS pMSSM dataset for our study.

In addition, we check the resulting SLHA files against SM and MSSM Higgs constraints using \texttt{HiggsSignals-1.4.0}~\cite{Bechtle:2013xfa,Stal:2013hwa,Bechtle:2013gu} and \texttt{HiggsBounds-4.3.1}~\cite{arXiv:0811.4169,arXiv:1102.1898,arXiv:1301.2345,arXiv:1311.0055,arXiv:1507.06706} respectively. HiggsBounds uses a number of experimental analyses~\cite{arXiv:1207.7214,arXiv:1011.1931,hep-ex/0206022,arxiv:1202.1488,arXiv:1008.3564,arXiv:1106.4782,arXiv:1108.3331,arxiv:1202.1997,arXiv:0707.0373,arXiv:1506.02301,arXiv:0806.0611,hep-ex/0111010,arXiv:1402.3051,arXiv:1107.1268,hep-ex/0501033,arXiv:1509.04670,arXiv:1012.0874,arXiv:1402.3244,arxiv:1112.2577,arXiv:0908.1811,arXiv:1307.5515,Abazov:2010ct,arxiv:1202.3478,hep-ex/0410017,arXiv:1404.1344,arXiv:1202.1416,arXiv:1406.7663,hep-ex/0107034,arXiv:1204.2760,arXiv:1504.00936,arXiv:1108.5064,arXiv:0906.1014,arXiv:1109.3357,arXiv:1202.1415,hep-ex/0107032,arXiv:1001.4468,arXiv:1109.3615,hep-ex/0401022,arXiv:1409.6064,hep-ex/0107031,arXiv:1207.0449,arXiv:1507.05930,arXiv:1106.4885,arXiv:0907.1269,arXiv:1207.6436,arXiv:1106.4555,arXiv:1003.3363,arXiv:0905.3381,arXiv:1406.5053,arXiv:1107.4960,arXiv:0809.3930,arXiv:1202.1414,hep-ex/0602042,arXiv:0906.5613,arxiv:1202.1408,arXiv:1407.6583,arXiv:1509.00389,hep-ex/0404012,arXiv:0901.1887,CDFnotes,D0notes,CMSnotes,ATLASnotes,LHWGnotes}, and internally uses several SM results~\cite{hep-ph/9704448,hep-ph/0102227,hep-ph/0102241,hep-ph/0201206,hep-ph/0207004,hep-ph/0302135,arXiv:0811.3458,Dawson:1990zj,Djouadi:1991tka,hep-ph/9504378,hep-ph/0404071,hep-ph/0407249,arXiv:0809.1301,arXiv:0809.3667,hep-ph/0306211,arXiv:0901.2427,hep-ph/0307206,hep-ph/0306234,hep-ph/0406152,hep-ph/0304035,hep-ph/9206246,hep-ph/9905386,hep-ph/0306109,hep-ph/0403194,hep-ph/0612172,hep-ph/0107081,hep-ph/0107101,hep-ph/0211438,hep-ph/0305321,arXiv:0705.2744,arXiv:0707.0381,arXiv:0710.4749,arXiv:1101.0593,arXiv:1201.3084,arXiv:1307.1347} to convert between experimental limits with different normalizations. As we aim to classify heavy Higgs decaying to SUSY particles, parameter sets where the neutral CP-even heavy Higgs solely decays to SM particles are removed.

The total dataset is divided into three categories, bino-like, wino-like and higgsino-like LSP,  according to the dominant part of the neutralino mixing matrix~$\mathcal{Z}$ (see table~\ref{tab_category}). It is important to realize that the nature of the neutralino signifies only the dominant part of the mixing matrix rather than a pure nature. This will be important while discussing the results. Our work focuses on the bino and higgsino dataset. Due to the small mass difference between the lightest chargino~$\tilde{\chi}_1^\pm$ and the wino-like LSP, a majority of the parameters in the wino-set result in long-lived charged particles, leading to displaced vertex signatures in the detector. Since the \texttt{SModelS} framework is not capable of handling displaced vertices, the wino-set is not processed in this work. None-the-less we will discuss salient features of Higgs decays to wino-like LSP. No additional constraints from the Higgs sector were applied to this dataset. In order to avoid any long-lived particles in the final states, we remove SLHA files containing particles with a decay length $c\tau>1$~mm from the bino and higgsino dataset as well.

\begin{table}[h]
\setlength\extrarowheight{5pt}
\centering
\begin{footnotesize}
\begin{tabular}{c c c}
\hline
\hline
\textbf{Type} & \textbf{Definition} & \textbf{Sample Size}\\[0.8ex]
\hline
Bino-like LSP & $\mathcal{Z}_{11}^2 > \text{max}(\mathcal{Z}_{12}^2, \mathcal{Z}_{13}^2 + \mathcal{Z}_{14}^2)$ & 31,112\\
Higgsino-like LSP & $(\mathcal{Z}_{13}^2 + \mathcal{Z}_{14}^2) > \text{max}(\mathcal{Z}_{11}^2, \mathcal{Z}_{12}^2)$ & 59,044\\
Wino-like LSP & $\mathcal{Z}_{12}^2 > \text{max}(\mathcal{Z}_{11}^2, \mathcal{Z}_{13}^2 + \mathcal{Z}_{14}^2)$ & 39,816\\[0.8ex]
\hline
\hline
\end{tabular}
\end{footnotesize}
\caption{Definition of the categorized parameter-sets~\cite{Aad:2015baa} and the sample size after applying all constraints. We do not apply additional constraints on the wino dataset and do not remove SLHA files containing long-lived charged particles from this set as mentioned in the text.}\label{tab_category}
\end{table}

The complete classification of the Higgs decay modes involves the dependence on the characteristics of each of the supersymmetric particle. It is hence difficult to work out all the combinations possible for the Higgs decay. We hence take help of \texttt{SModelS} which can automatically compute all possible decay modes and reports the output in a convenient format which can be interpreted further. 
\subsection*{SModelS Framework}
\noindent
\texttt{SModelS}~\cite{Ambrogi:2017neo,Kraml:2013mwa,SModelS:wiki} decomposes an input spectrum into Simplified Model Spectrum (SMS) topologies to test a given BSM scenario against SMS limits from LHC direct SUSY searches. The code can handle either a Les Houches Event (LHE) file or an SLHA file as input. Here, we use SLHA input. The code further uses \texttt{PySLHA}~\cite{Buckley:2013jua} to read the input files.

Given an input file, containing production cross section and branching ratios of $\mathbb{Z}_2$-odd BSM particles, the code first constructs all possible combinations of production cross sections $\sigma_{odd}$ times branching ratio $\mathcal{B}$. It is important to know that such combinations of $(\sigma_{odd}\times\mathcal{B})_i$, called elements, depend on the underlying parameter space and hence different SLHA files will lead to different elements. For each element, \texttt{SModelS} stores information on the masses of BSM particles involved and the theory predictions $(\sigma_{odd}\times\mathcal{B})_i$. This feature of dynamically constructing all possible elements is crucial to identify simplified decay chains as heavy Higgs decays can lead to different decay chains, which is difficult to classify analytically. 

The elements which result from decomposition are then grouped together depending on the required information. For testing against the experimental results, elements are grouped according to assumptions on experimental results\footnote{For completeness, electrons and muons are combined to form light leptons (labeled `l'), as well as gluons and light quarks (u, d, c, s) are combined to `$jets$'.}. These can include, the mass vector of BSM particles involved and the charges or flavor of SM final states. To identify elements which are not currently tested by experimental results, all elements leading to the same final state are summed over, irrespective of the BSM mass vector. This is the so-called missing topologies module, which will be heavily used in this study. The missing topologies are accompanied by their theoretical prediction of $(\sigma_{odd}\times\mathcal{B})_i$. 

Finally, it is difficult to form simplified models with elements resulting in  long decay chains. The code hence provides information about the mother particles and the theoretical prediction of $\sigma_{odd}\times\mathcal{B}$ of decay chains with more than one intermediate state in at least one branch. The theoretical prediction of $\sigma_{odd}\times\mathcal{B}$ includes all final states with same initially produced mother particles. The long cascade decays thus symbolize complicated topologies where it might not be possible to define simplified models. The weights $(\sigma_{odd}\times\mathcal{B})_i$ of these decay chains also give an estimate of the signal cross sections, which cannot be constrained via simple interpretations of experimental searches.

To reduce the computing time of the decomposition, a so-called $\sigma_{cut}$ parameter is used in \texttt{SModelS}. It allows to neglect decay chains with cross sections times branching ratio below $\sigma_{cut}$. Since we use \texttt{SModelS} for the calculation of a rather low production cross section of the mother particle (here heavy Higgs) and aim to classify all possible decay modes, we thus use an equally low $\sigma_{cut} = 10^{-9}$~fb. Despite the very low $\sigma_{cut}$, the final signal cross sections in a given final state can be significant, as \texttt{SModelS} sums over individual elements adding the corresponding $(\sigma_{odd}\times \mathcal{B})_i$. It is worth noting that an optimal use of \texttt{SModelS} can be achieved by setting the $\sigma_{cut}$ parameter to a relative value compared to the BSM production cross section. This is particularly important for using \texttt{SModelS} for analyzing BSM models with very low production cross sections.

When the mass difference between two BSM states ($\Delta m$) is small, a mass compression is performed by \texttt{SModelS}. Mass compression effectively removes the decay chain which results in soft objects at the detector and replaces the LSP by an effective LSP mass. The main advantage of performing a mass compression is that the resulting simplified topology is shorter than the original one and soft final states are reduced. We set $\Delta m = 5$~GeV for this procedure. This is of particular importance for the higgsino dataset where the mass difference between the two lightest neutralinos and the lightest chargino is small. Thus, a decay of the latter can be undetected.

Furthermore, invisible compression is used to disregard invisible decay products occurring at the end of  decay chains in the SMS topologies. 

During the decomposition procedure and in the final output, the \texttt{SModelS} framework uses bracket notation describing the simplified model topologies. However, for better readability we use a simplified version of the bracket notation with every branch being enclosed by parenthesis and vertices being separated with commas. For simplicity, we write only the SM final states and do not label the intermediate BSM states. Existence of MET in each of the branch at the end of decay chain is assumed and we do not explicitly write it. For example, $p p \rightarrow \tilde t_1 \bar{\tilde t}_1$, with $\tilde t_1\rightarrow \tilde\chi^0_1 t$, the so called T2tt topology, will be written as [[[t]],[[t]]] within \texttt{SModelS}, while we will represent it as (t),(t). 

Currently, \texttt{SModelS} is applicable for any BSM scenario obeying a $\mathbb{Z}_2$ symmetry and resulting in a MET final state at the LHC. The formalism is thus not capable of directly decomposing the heavy Higgs decays to supersymmetric particles, as it does not obey $\mathbb{Z}_2$ symmetry. Nevertheless, we demonstrate that it is possible to use \texttt{SModelS} for resonant heavy Higgs searches in computing the MSSM Higgs production cross section using \texttt{SusHi-1.6.1}~\cite{Harlander:2012pb, Harlander:2016hcx}. Equipped with these cross sections, we calculate the product of $\sigma$ and the branching ratio  to SUSY particles, thus effectively constructing 
$\sigma_{odd}$ necessary as an input to \texttt{SModelS}. This is a particularly new and innovative way to using \texttt{SModelS} for resonant searches.

An important change in our procedure is the reduction in the results database. Our primary aim is to classify all important heavy Higgs decays to SUSY particles. By definition, missing topologies in \texttt{SModelS} are topologies currently not present in the \texttt{SModelS} experimental results database. We hence artificially reduce the database in order to globally classify all heavy Higgs decays within the the missing topology module. For the results presented below we have only kept the T2tt topology in the \texttt{SModelS} database\footnote{Using \texttt{SModelS} without a database is not possible, which is why we artificially reduce it to include only one insignificant topology.}, as the stops in the pMSSM dataset are in general heavy and a decay into a stop pair would, if kinematically allowed, not result in large signal. This has been checked numerically.

\section{Results}
\label{sec:results}
\begin{figure}[h]
\centering
\includegraphics[trim={0cm 4.5cm 0cm 0.8cm}, clip, width=\textwidth]{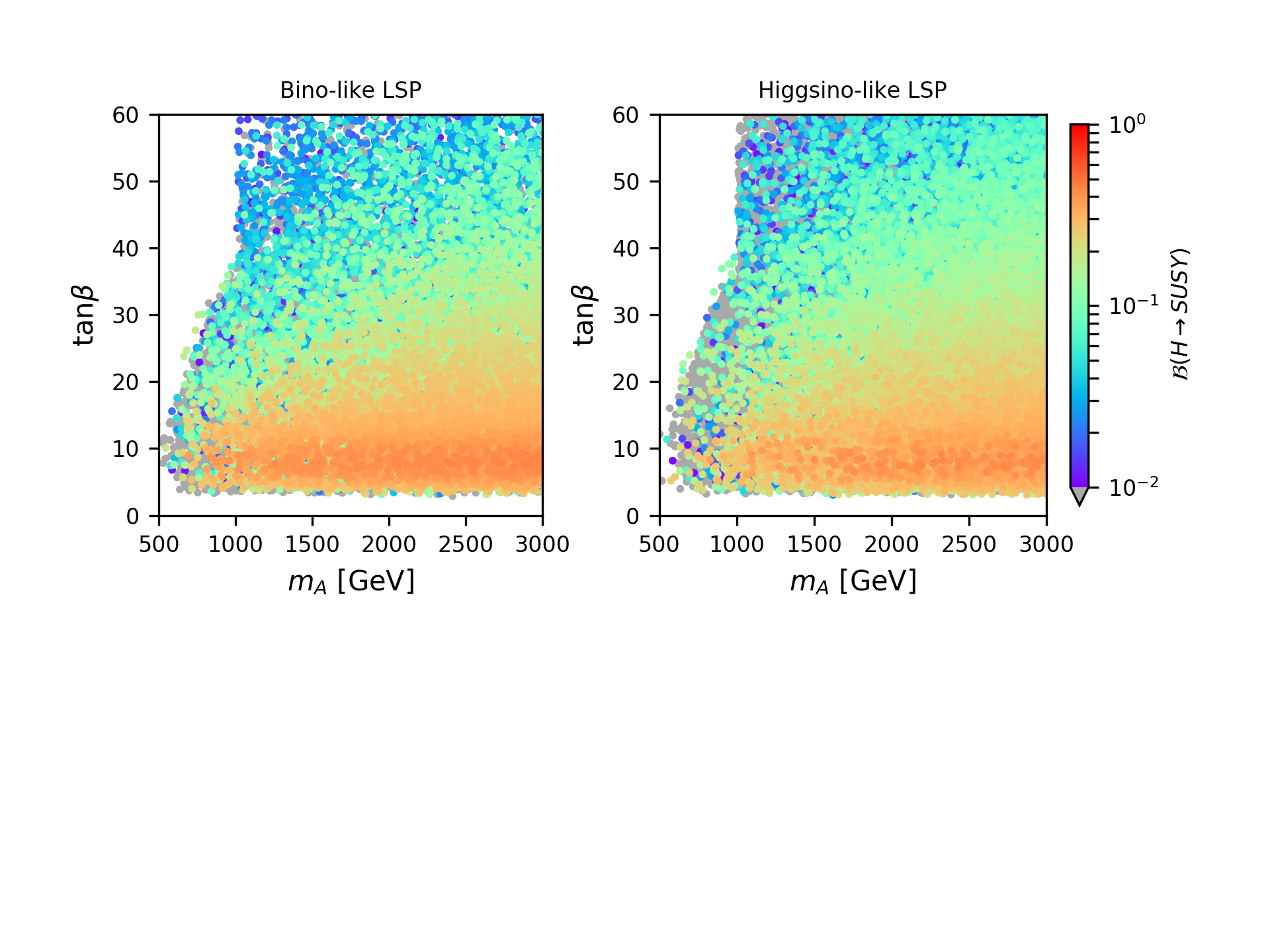}
\caption{Branching ratio of heavy Higgs bosons decaying to SUSY particles in the $m_A$-$\tan\beta$ parameter plane for bino-like LSPs (left) and higgsino-like LSPs (right).}\label{fig_BR_3D}
\end{figure}

We first understand the importance of supersymmetric decays of a heavy Higgs in our dataset. Previous works already showed a potentially high branching ratio of $H$ to weakinos for moderate $\tan\beta$ values~\cite{Barman:2016kgt}. We study the branching ratio of $H$ to all SUSY particles for the full 19-dimensional pMSSM parameter space of the ATLAS pMSSM dataset.  Fig.~\ref{fig_BR_3D} shows a peak branching ratio of about $\mathcal{B}(H\rightarrow SUSY) \approx 60\%$ for $\tan\beta < 15$. 

In addition to decays to weakinos, it is also possible that Higgs decays to other sparticles, all contributions are taken into account in the plot. However, the main contribution comes from decays to weakinos, which has been checked numerically. The branching ratio value is decreasing to about $10\%$ for high $\tan\beta$ and is nearly independent of $m_A$. However, the approximately constant $\sigma\times\mathcal{B}$ in dependence of $\tan\beta$ seen later is the result of increasing heavy Higgs production cross sections for regions with decreasing branching ratios. For large $\tan\beta$ values, the enhanced coupling of heavy Higgs bosons to down-type quarks leads to bottom-quark annihilations becoming the dominant production process.

Current LHC searches are effective for SM decays of heavy Higgs bosons and miss out on the parameter space with large branching ratios to SUSY particles. It is possible to reinterpret existing LHC SUSY searches in order to cover some of this parameter space~\cite{Bisset:2007mi, Barman:2016kgt}. Therefore, a sole analysis of $H\rightarrow SM$ is insufficient to cover the entire parameter space for heavy Higgs and it becomes interesting to ask which final states should the experiments consider to successfully cover the whole parameter space. 

With this aim, we analyze the pMSSM dataset using the \texttt{SModelS} setup as described in the previous section. The analysis of the missing topologies can be presented in two ways, either by showing the most frequently occurring missing topologies, or by selecting for each parameter point the missing topology with the highest cross section, the so-called `most dominant missing topology'.

Before we quantify our results, it is important to realize two features of Higgs decays to weakinos. As the Higgs boson couples to the admixture of gaugino-higgsino states, asymmetric decays to weakinos are preferred over symmetric decays if kinematically allowed. For the higgsino dataset, the $\tilde{\chi}^0_1, \tilde{\chi}^0_2, \tilde{\chi}^{\pm}_1$ are higgsino-like and heavier states are gaugino-like. For the bino dataset, $\tilde{\chi}^0_2$ can either be wino-like or higgsino-like. If $\tilde{\chi}^0_2$ is wino-like, the decays of Higgs to $\tilde{\chi}^0_2 \tilde{\chi}^0_1$ are suppressed and decays via heavier weakino states are preferred. It should be noted, that depending on the characteristics of the weakino, thus the ordering of the underlying parameters, the heavy Higgs can prefer decays to a heavier weakino state, rather than the next-to-LSP (NLSP) if kinematically allowed. Secondly, when decays to heavier states are kinematically not allowed, decays to lighter states will be preferred due to phase space even when the couplings are suppressed. This particularly matters for a light heavy Higgs ($m_H \lesssim 1.5$~TeV) within the reach of LHC14.

\begin{figure}[h]
\centering
\includegraphics[trim={0cm 4.5cm 0cm 1cm},clip,width=\textwidth]{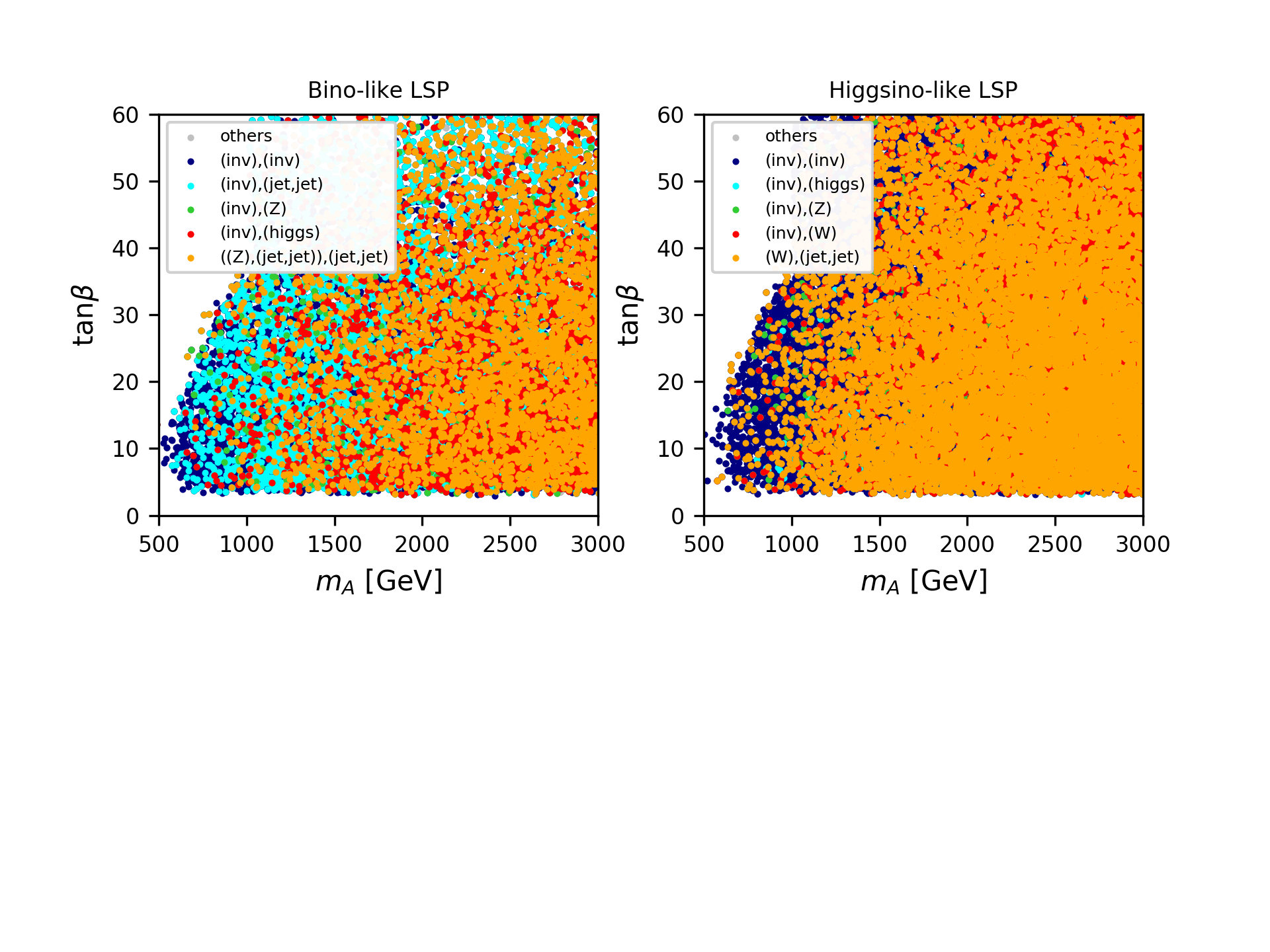}
\caption{Most frequently occurring missing topologies in the $m_A$-$\tan\beta$ parameter plane for bino-like LSPs (left) and higgsino-like LSPs (right). The topologies are dominated by mono-$X$ signatures with $X$ being on- and off-shell $W$, $Z$ or the light Higgs~$h$.}\label{fig_freq}
\end{figure}

With this knowledge, in fig.~\ref{fig_freq}, we show the five most frequently occurring topologies for both the bino- (left panel) and higgsino-like (right panel) dataset in the $m_A$-$\tan\beta$ plane. The absence of any points below $m_A \lesssim 500$ GeV is due to direct Higgs searches at the LHC. Thus, the analyzed data includes only decoupled Higgs states leading to the coupling of the heavy Higgs bosons to vector bosons being suppressed.

Followed by invisible final states, in either datasets, the most prominent missing topologies are the mono-$X$ signatures ($X$ = $h$, $W$, $Z$). They occur as a result of the asymmetric decay of a heavy Higgs to a pair of neutralinos or charginos, e.g. a pair of $\tilde\chi^0_i\tilde\chi^0_j$ ($i \neq j$). In mono-$X$ topologies, the lighter state then decays softly to the LSP.

The most frequent topology for both, bino- and higgsino-like LSPs, with the highest frequency is (inv),(inv). This signature is the result of either extreme mass splitting between LSP and heavy weakinos, or nearly degenerate weakino states. The former leads to $H$ decaying to a pair of LSPs, where this process is the only kinematically allowed decay mode of $H$ to SUSY particles. However, as the heavy Higgs prefers asymmetric decays, the latter leads to large $\sigma\times\mathcal{B}$. There, $H$ decays to the LSP and a light neutralino (e.g. $\tilde{\chi}_2^0$) or a pair of light charginos ($\tilde{\chi}_1^\pm$), where heavier states have to decay softly. It is worth noting that (inv),(inv) occurs for the lowest Higgs mass $m_H \sim 500 \,\rm{GeV}$. The occurrence of (inv),(inv) at low Higgs mass is due to kinematics, where decays to higher weakinos are forbidden, this also makes it the most frequently occurring topology. 

The next topology for higgsino-like LSPs is mono-$W$. This signature is mainly the result of $H$ decays to a pair of charginos ($H\rightarrow\tilde{\chi}_1^\pm\tilde{\chi}_2^\mp$). For the higgsino-like dataset, the chargino decays invisibly due to a small mass splitting between chargino and LSP. However in limited cases, the latter can be visible to the detector, leading to (W),(jet,jet) signatures. Mono-$W$ signatures can also be the result of the Higgs decaying to a pair of neutralinos  ($\tilde\chi^0_2\tilde\chi^0_1$), where the heavier neutralino state decays to a $W$ boson and the lightest chargino. The latter then decays softly.

Only a minority of the bino-like LSP dataset contains mono-$W$ signatures, as the heavy chargino state can be significantly heavier than the LSP, thus an asymmetric Higgs decay to charginos is kinematically forbidden or the additional decays of the heavy chargino lead to more complex topologies. For a bino-like LSP, we instead find a mono-$Z$ topology, which occurs when the Higgs decays to pair of neutralinos. 

Furthermore, we find mono-$X$ signatures with off-shell $X$, where for off-shell states, the dominating channels are $Z^*,W^*\rightarrow jet\,jet$ and $Z^*,h^*\rightarrow b\bar{b}$. In a limited number of files,  the soft $W$ boson decays leptonically ($W^*\rightarrow l\,\nu_l$), where a single $1^\text{st}$ or $2^\text{nd}$ generation lepton ($l \in \left\lbrace e,\mu\right\rbrace$) is visible to the detector. Signatures with additional on- and off-shell vector bosons are the result of the Higgs decay to heavier neutralinos. This special case occurs more frequently in the bino-like LSP dataset, as addressed later.

The impact of kinematics on each of the topology is also clearly visible in the plots. The invisible final state and off-shell decays occur at low Higgs mass, while the mono-$X$ topologies, which require substantial mass difference between LSP and NLSP correspondingly occur at higher masses $m_H\sim 1\, \rm{TeV}$. 

It has been pointed out that simplified models resulting in mono-$W$/$Z$ final states often run into unitarity violations unless carefully constructed~\cite{Haisch:2016usn}. In this light, these results demonstrating the importance of mono-$X$ topologies are interesting. Here we obtain a theoretically consistent theory parameter space which does not violate unitarity.

\begin{figure}[h]
\centering
\includegraphics[trim={0cm 4.5cm 0cm 1cm},clip,width=\textwidth]{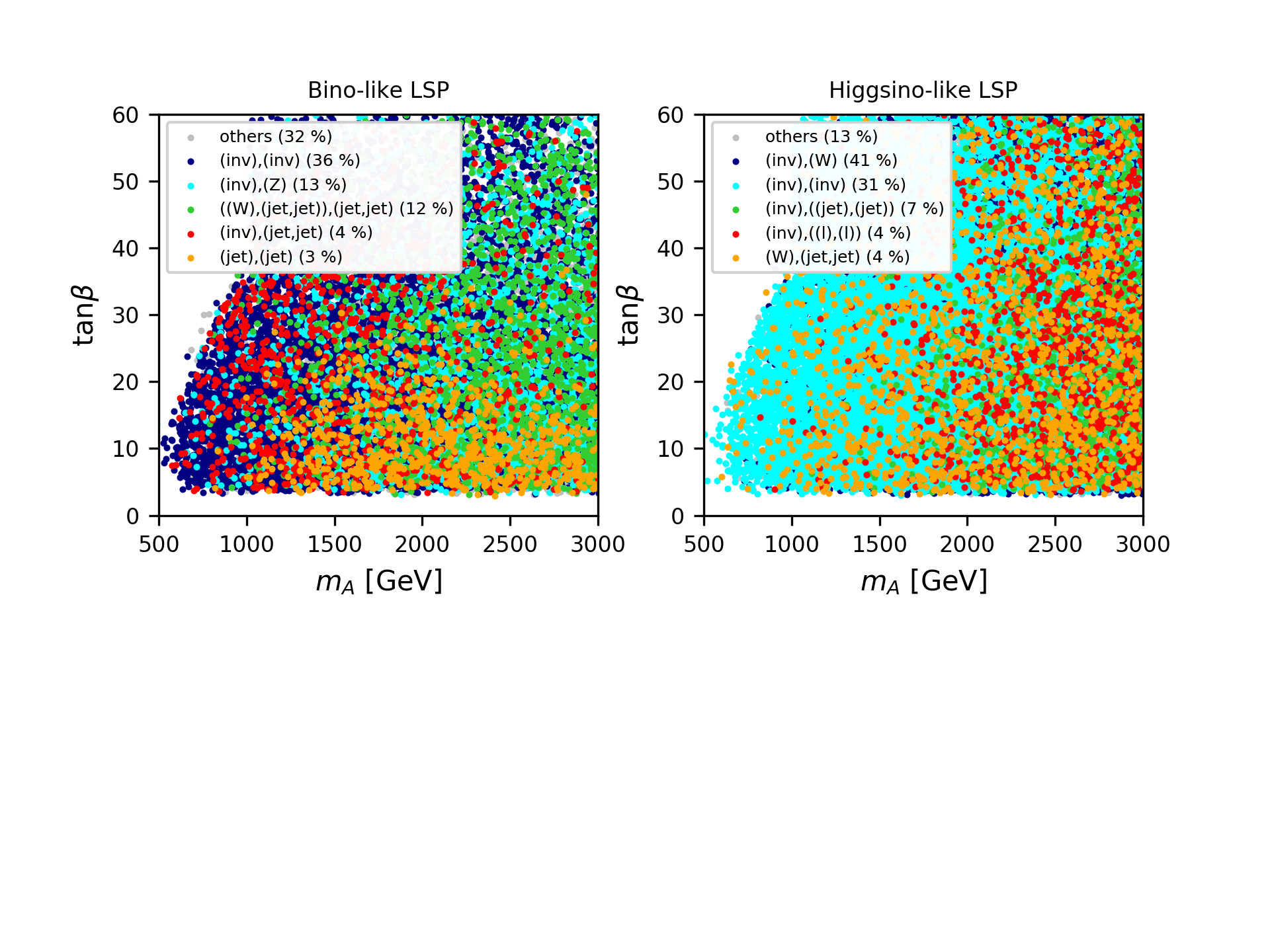}
\caption{Most dominating missing topologies in the $m_A$-$\tan\beta$ parameter plane for bino-like LSPs (left) and higgsino-like LSPs (right). The percent value gives the fraction of the dataset, where this topology has the highest $\sigma\times\mathcal{B}$ value.}\label{fig_dom}
\end{figure}

While the missing topologies sorted on their frequencies demonstrate the importance of mono-$X$ topologies, it is unclear whether they are also the topologies with highest cross section at each point. In order to clarify this point, in fig.~\ref{fig_dom}, we plot topologies selecting for each parameter point the missing topology with the highest cross section. A different picture emerges in this case.

For the higgsino-like LSP dataset the preference to asymmetric decays leads to the largest cross sections for small values of the wino- and higgsino mass parameter, $M_2$ and $\mu$, respectively. Thus, we find preferred decays to wino-/higgsino-like neutralinos ($\tilde{\chi}_i^0$, $i\in \lbrace 1,2,3\rbrace$) or to a pair of charginos, where the latter results in the highest cross sections. The large branching ratio of the Higgs to charginos and the fact that \texttt{SModelS} sums over all mother particles leaving the same signature, leads to mono-$W$ being the dominating topology.

For bino-like LSPs, the results are less obvious. Preferred decays of Higgs to gaugino-/higgsino-mixtures lead to the highest cross sections for highly mixed bino-/higgsino-like LSPs, thus small values of the $M_1$ and $\mu$ parameters. Compared to the higgsino-like LSP dataset, these parameter sets have larger mass differences between the LSP and the higgsino-like chargino or NLSP. Furthermore, the wino-like $\tilde{\chi}_2^\pm$ and $\tilde{\chi}_4^0$ can be significantly heavier than the LSP. Thus, asymmetric decays of the Higgs are more likely to be kinematically forbidden, leading to a dominant invisible signature (inv),(inv). Nevertheless, the wino-like weakinos ($\tilde{\chi}_4^0$, $\tilde{\chi}_2^\pm$) play a crucial role in this dataset. If kinematically allowed, a decay of the Higgs to higgsino- and wino-like weakinos ($H\rightarrow \tilde{\chi}_i^0\tilde{\chi}_4^0$, with $i\in\lbrace 2,3 \rbrace$, and $H\rightarrow \tilde{\chi}_1^\pm\tilde{\chi}_2^\mp$) has large branching ratios. This leads to more complex topologies with on- and off-shell vector bosons, e.g. ((W/Z),(jet,jet)),(jet,jet).

Even though for the bino-like LSP dataset, the vertex $\tilde{\chi}_i^0\rightarrow h\tilde{\chi}_1^0$ ($i \in \lbrace 2,3\rbrace$) is larger compared to $\tilde{\chi}_i^0\rightarrow Z\tilde{\chi}_1^0$, we find mono-$Z$ signatures dominating the Higgs decay. This is the result of more possible decay modes for the decay $\tilde{\chi}_i\rightarrow Z\tilde{\chi}_j$, as e.g. the mass difference of two neutralinos being below the light Higgs mass. \texttt{SModelS} sums over the initial states with same final signatures, leading to larger $\sigma\times\mathcal{B}$ values for mono-$Z$ signatures.

An important difference between most frequent and most dominant topologies is the reduced importance of mono-$X$ topologies and emergence of more complicated decay chains for the latter. Given the complex structure of the weakino sector and the dependence on multiple soft parameters, it is not unexpected to see such complicated decay chains emerging. It is also interesting to note that most dominant topologies involve decays of weakinos via intermediate sleptons or squarks. Despite intense searches, it is possible for first two generation squarks/sleptons to be light. The direct decays of the Higgs are suppressed by the small Yukawa coupling, however it is possible that heavier weakinos decay to squarks/sleptons and this is precisely seen in the figure. Such topologies however occur for rather heavy Higgs masses $m_H \gtrsim 1\,\rm{TeV}$ and only a small number of points contain these signatures.
\begin{table}[h!]
\setlength\extrarowheight{5pt}
\centering
\begin{footnotesize}
\begin{tabular}{l l l}
\hline
\hline
\textbf{Topology}\hspace{1cm} & \textbf{Decay Processes}\hspace{5.5cm} & \textbf{Information}\\[.8ex]
\hline
(inv),(inv) & $H \rightarrow \tilde{\chi}_k^\pm \, \tilde{\chi}_l^\mp \rightarrow W^*_\text{soft} \, \tilde{\chi}_1^0 \, W^*_\text{soft} \, \tilde{\chi}_1^0$ & $k,l \in \left\lbrace 1,2\right\rbrace$\\
& $H \rightarrow \tilde{\chi}_1^0 \, \tilde{\chi}_1^0$ & $i,j \in \left\lbrace 1,2,3,4\right\rbrace$\\
& $H \rightarrow \tilde{\chi}_i^0 \, \tilde{\chi}_j^0 \rightarrow X^*_\text{soft} \, \tilde{\chi}_1^0 \, X'^*_\text{soft} \, \tilde{\chi}_1^0$ & $X,X' \in \left\lbrace Z,h\right\rbrace$\\[0.8ex]
\hline
(inv),(higgs) & $H \rightarrow \tilde{\chi}_1^0 \, \tilde{\chi}_j^0 \rightarrow \tilde{\chi}_1^0 \, h \, \tilde{\chi}_1^0$ & $j \in \left\lbrace 2,3,4\right\rbrace$\\
& $H \rightarrow \tilde{\chi}_1^\pm \, \tilde{\chi}_2^\mp \rightarrow W^*_\text{soft} \, \tilde{\chi}_1^0 \, h \, \tilde{\chi}_1^\mp \rightarrow W^*_\text{soft} \, \tilde{\chi}_1^0 \, Z \, W^*_\text{soft} \, \tilde{\chi}_1^0$ & \\[0.8ex]
\hline
(inv),(Z) & $H \rightarrow \tilde{\chi}_1^0 \, \tilde{\chi}_j^0 \rightarrow \tilde{\chi}_1^0 \, Z \, \tilde{\chi}_1^0$ & $j \in \left\lbrace 2,3,4\right\rbrace$\\
& $H \rightarrow \tilde{\chi}_1^\pm \, \tilde{\chi}_2^\mp \rightarrow W^*_\text{soft} \, \tilde{\chi}_1^0 \, Z \, \tilde{\chi}_1^\mp \rightarrow W^*_\text{soft} \, \tilde{\chi}_1^0 \, Z \, W^*_\text{soft} \, \tilde{\chi}_1^0$ & \\[0.8ex]
\hline
(inv),(W) & $H \rightarrow \tilde{\chi}_1^\pm \, \tilde{\chi}_2^\mp \rightarrow W^*_\text{soft} \, \tilde{\chi}_1^0 \, W \, \tilde{\chi}_1^0$ & $j \in \left\lbrace 3,4\right\rbrace$\\
 & $H \rightarrow \tilde{\chi}_1^0 \, \tilde{\chi}_j^0 \rightarrow \tilde{\chi}_1^0 \, W \, \tilde{\chi}_1^\pm \rightarrow \tilde{\chi}_1^0 \, W \, W^*_\text{soft} \, \tilde{\chi}_1^0$ &\\[0.8ex]
\hline
(inv),(jet,jet) & $H \rightarrow \tilde{\chi}_1^0 \, \tilde{\chi}_j^0 \rightarrow \tilde{\chi}_1^0 \, Z^* \, \tilde{\chi}_1^0$ & $j \in \left\lbrace 2,3,4\right\rbrace$\\
& $H \rightarrow \tilde{\chi}_1^\pm \, \tilde{\chi}_2^\mp \rightarrow W^*_\text{soft} \, \tilde{\chi}_1^0 \, W^* \, \tilde{\chi}_1^0$ &\\
& $H \rightarrow \tilde{\chi}_1^\pm \, \tilde{\chi}_2^\mp \rightarrow W^*_\text{soft} \, \tilde{\chi}_1^0 \, Z^* \, \tilde{\chi}_1^\mp \rightarrow W^*_\text{soft} \, \tilde{\chi}_1^0 \, Z^* \, W^*_\text{soft} \, \tilde{\chi}_1^0 $ &\\[0.8ex]
\hline
(W),(jet,jet) & $H \rightarrow \tilde{\chi}_1^\pm \, \tilde{\chi}_2^\mp \rightarrow W^* \, \tilde{\chi}_1^0 \, W \, \tilde{\chi}_1^0$ & $i,j \in \left\lbrace 2,3,4\right\rbrace,$\\
& $H \rightarrow \tilde{\chi}_i^0 \, \tilde{\chi}_j^0 \rightarrow Z^* \, \tilde{\chi}_1^0 \, W \, \tilde{\chi}_1^\pm \rightarrow Z^* \, \tilde{\chi}_1^0 \, W \, W_\text{soft}^* \, \tilde{\chi}_1^0$ & $i<j$\\[0.8ex]
\hline
((Z),(jet, jet)),(jet,jet) & $H \rightarrow \tilde{\chi}_1^\pm \, \tilde{\chi}_2^\mp \rightarrow W^* \, \tilde{\chi}_1^0 \, Z \, \tilde{\chi}_1^\mp \rightarrow W^* \, \tilde{\chi}_1^0 \, Z \, W^* \, \tilde{\chi}_1^0$ & $i,j,k \in \left\lbrace 2,3,4\right\rbrace,$\\
& $H \rightarrow \tilde{\chi}_i^0 \, \tilde{\chi}_j^0 \rightarrow Z^* \, \tilde{\chi}_1^0 \, Z \, \tilde{\chi}_k^0 \rightarrow Z^* \, \tilde{\chi}_1^0 \, Z \, Z^* \, \tilde{\chi}_1^0$ & $i,k<j$\\[0.8ex]
\hline
((W),(jet, jet)),(jet,jet) & $H \rightarrow \tilde{\chi}_1^\pm \, \tilde{\chi}_2^\mp \rightarrow W^* \, \tilde{\chi}_1^0 \, W \, \tilde{\chi}_j^0 \rightarrow W^* \, \tilde{\chi}_1^0 \, W \, Z^* \, \tilde{\chi}_1^0$ & $i,j \in \left\lbrace 2,3,4\right\rbrace,\,i<j$\\
& $H \rightarrow \tilde{\chi}_i^0 \, \tilde{\chi}_j^0 \rightarrow Z^* \, \tilde{\chi}_1^0 \, W \, \tilde{\chi}_k^\pm \rightarrow Z^* \, \tilde{\chi}_1^0 \, W \, W^* \, \tilde{\chi}_1^0$ & $k \in \left\lbrace 1,2\right\rbrace$\\[0.8ex]
\hline
(inv),((jet),(jet)) & $H \rightarrow \tilde{\chi}_1^0 \, \tilde{\chi}_j^0 \rightarrow \tilde{\chi}_1^0 \, jet \, \tilde{q} \rightarrow \tilde{\chi}_1^0 \, jet \, jet \, \tilde{\chi}_1^0$ & $j \in \left\lbrace 2,3,4\right\rbrace$\\
& $H \rightarrow \tilde{\chi}_1^\pm \, \tilde{\chi}_2^\mp \rightarrow W_\text{soft}^* \, \tilde{\chi}_1^0 \, jet \, \tilde{q} \rightarrow W_\text{soft}^* \, \tilde{\chi}_1^0 \, jet \, jet \, \tilde{\chi}_1^0$ & $\tilde{q} \in \left\lbrace \tilde{u},\tilde{d},\tilde{c},\tilde{s}\right\rbrace_{L,R}$\\[0.8ex]
\hline
(inv),((l),(l)) & $H \rightarrow \tilde{\chi}_1^0 \, \tilde{\chi}_j^0 \rightarrow \tilde{\chi}_1^0 \, l \, \tilde{l} \rightarrow \tilde{\chi}_1^0 \, l \, l \, \tilde{\chi}_1^0$ & $j \in \left\lbrace 2,3,4\right\rbrace$\\
& $H \rightarrow \tilde{\chi}_1^\pm \, \tilde{\chi}_2^\mp \rightarrow W_\text{soft}^* \, \tilde{\chi}_1^0 \, l' \, \tilde{l} \rightarrow W_\text{soft}^* \, \tilde{\chi}_1^0 \, l' \, l \, \tilde{\chi}_1^0$ & $\tilde{l} \in \left\lbrace \tilde{e},\tilde{\mu}\right\rbrace_{L,R}$\\
& $H \rightarrow \tilde{\chi}_1^0 \, \tilde{\chi}_j^0 \rightarrow \tilde{\chi}_1^0 \, W^* \, \tilde{\chi}_1^\pm \rightarrow \tilde{\chi}_1^0 \, W^* \, W^* \, \tilde{\chi}_1^0$ & $W^*\rightarrow l\,\nu_l$\\[0.8ex]
\hline
(jet),(jet) & $H \rightarrow \tilde{q} \, \tilde{q} \rightarrow jet \, \tilde{\chi}_1^0 \, jet \, \tilde{\chi}_1^0$ & $\tilde{q} \in \left\lbrace \tilde{u},\tilde{d},\tilde{c},\tilde{s}\right\rbrace_{L,R}$\\
& $H \rightarrow \tilde{q} \, \tilde{q} \rightarrow jet \, \tilde{\chi}_k^\pm \, jet \, \tilde{\chi}_k^\pm \rightarrow jet \, W^*_\text{soft} \, \tilde{\chi}_1^0 \, jet \, W^*_\text{soft} \, \tilde{\chi}_1^0$ & $k \in \left\lbrace 1,2 \right\rbrace$ \\[0.8ex]
\hline
(l),(l) & $H \rightarrow \tilde{l} \, \tilde{l} \rightarrow l \, \tilde{\chi}_1^0 \, l \, \tilde{\chi}_1^0$ & $\tilde{l} \in \left\lbrace \tilde{e},\tilde{\mu}\right\rbrace_{L,R}$\\
& $H \rightarrow \tilde{l} \, \tilde{l} \rightarrow l \, \tilde{\chi}_k^\pm \, l \, \tilde{\chi}_k^\pm \rightarrow l \, W^*_\text{soft} \, \tilde{\chi}_1^0 \, l \, W^*_\text{soft} \, \tilde{\chi}_1^0$ & $k \in \left\lbrace 1,2 \right\rbrace$ \\[0.8ex]
\hline
\hline
\end{tabular}
\end{footnotesize}
\caption{Missing topologies for heavy Higgs decays and their main possible decay processes given by the \texttt{SModelS} output. The subscript `soft' represents undetected soft decay products ($m_\text{soft}<5$~GeV) and `*' notes off-shell particles. Electrons and muons are combined to light leptons ($l$) and gluons and light quarks ($u$, $d$, $c$, $s$) are combined to $jets$. It should be noted that LSPs as decay products can be replaced by the process $\tilde{\chi} \rightarrow X^* _\text{soft} \, \tilde{\chi}_1^0$, where $\tilde{\chi}$ can be any neutralino or chargino and $X^* _\text{soft}$ can be any non-colored state not visible to the detector. }\label{tab_decay}
\end{table}

Although it is possible to have light first two generation squarks/sleptons, lower bounds on the stop and sbottom masses disfavor a direct decay of the heavy Higgs to either ($H\rightarrow\tilde{t}\tilde{t}/\tilde{b}\tilde{b}$). The LHC limits also forbid stops and sbottoms as an intermediate state. It is worth noting that, for bino-like LSPs we find dijet and dilepton signatures (not shown in the plots) as well. Nevertheless, dijet and dilepton signatures are insignificant in the total cross section.

The detailed physical processes underlying the missing topologies is given in table~\ref{tab_decay}. It should be noted, that depending on the characteristics of the weakinos and thus the hierarchy of the soft parameters, the decay to two heavier states (e.g. $H\rightarrow \tilde{\chi}^0_3\tilde{\chi}^0_2$) can be preferred compared to e.g. $H\rightarrow \tilde{\chi}^0_3\tilde{\chi}^0_1$. Thus, in the table, the LSP can be replaced by any other non-colored state, which then decays softly. Additionally, more complex decays have to be considered, as \texttt{SModelS} sums over all mother particles with the same final signature.

\subsection*{Mono-$X$ Final States}
\begin{figure}[h]
\centering
\includegraphics[trim={0cm 4.5cm 0cm 0.8cm},clip,width=\textwidth]{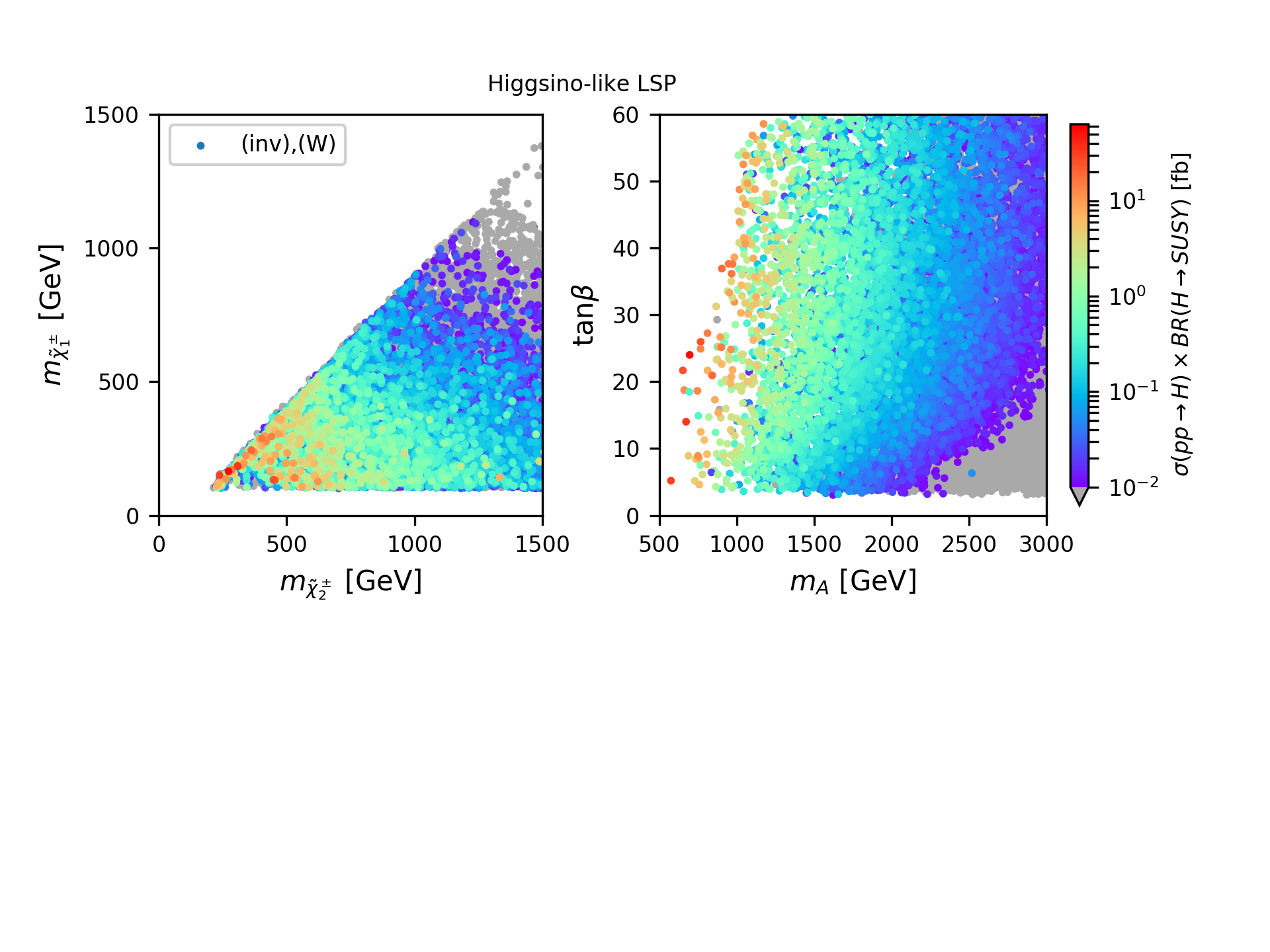}
\caption{Cross sections for heavy Higgs searches with mono-$W$ signature in the $\tilde{\chi}^\pm_2$-$\tilde{\chi}^\pm_1$ mass-plane (left) and the $m_A$-$\tan\beta$ parameter plane (right) for higgsino-like LSPs.}\label{fig_monow}
\end{figure}

In this section, we discuss mono-$X$ signatures ($X$ = $h$, $W$, $Z$) in more detail. While the previous discussion signifies that mono-$X$ topologies would be very important to explore Higgs decays to SUSY particles, it gives no indication of the strength of the signal. To address this question, in fig.~\ref{fig_monow} we show the cross section for the mono-$W$ topology in the relevant weakino mass plane (left panel), and in the $m_A$-$\tan\beta$ parameter plane (right panel). It is worth noting that we obtain cross sections up to $10$~fb for $\sqrt{s}=14$~TeV. We retrieve similar cross sections for the other mono-$X$ ($X$ = $h$, $Z$) topologies in the higgsino-like and bino-like LSP dataset, but we do not show them here.

In each of the cases, a high cross section for the mono-$X$ topology can be obtained for a light MSSM Higgs mass (here: $m_A\approx m_H$), it falls smoothly as the mass of the Higgs increases as expected. The nearly constant behavior of $\sigma\times\mathcal{B}$ with $\tan\beta$ is the result of a strongly decreasing branching ratio being compensated by an increasing total production cross section of the heavy Higgs boson. The latter is the result of an enhanced Yukawa coupling of Higgs to down-type quarks for large $\tan\beta$, thus bottom-quark annihilation dominating the Higgs production. It is interesting to note that the mass difference in the LSP and NLSP can be large giving rise to potentially hard $X$ in the final state. 

\begin{figure}[h]
\centering
\includegraphics[trim={0cm 4.5cm 0cm 0.8cm},clip,width=\textwidth]{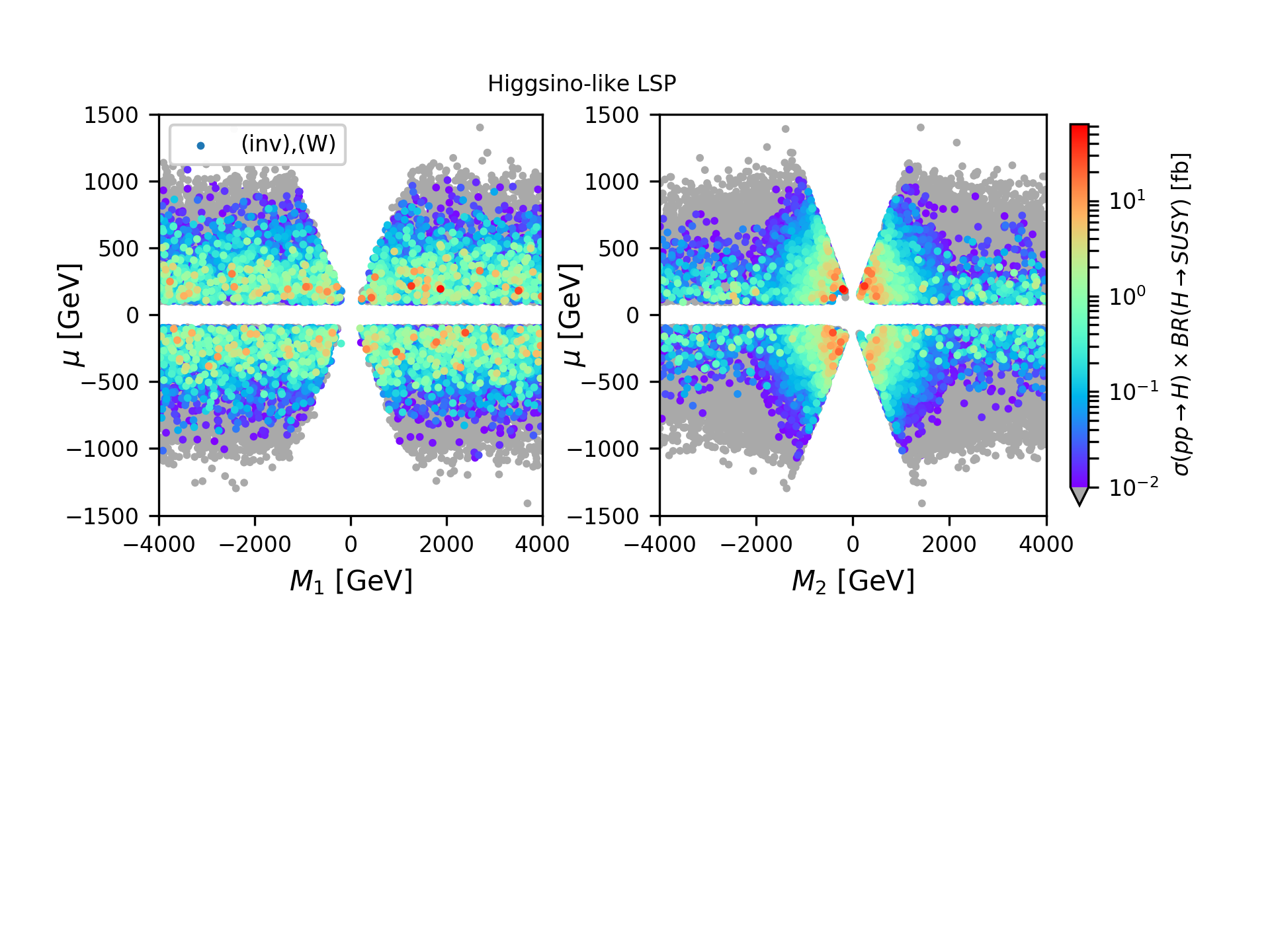}
\caption{Cross sections for heavy Higgs searches with mono-$W$ signature in the $M_1$-$\mu$ (left) and the $M_2$-$\mu$ soft parameter plane (right) for higgsino-like LSPs.}\label{fig_para_3D_higgsino}
\end{figure}

Furthermore, we study the dependence of the resulting cross sections on the underlying parameters. Fig.~\ref{fig_para_3D_higgsino} shows the production cross section of heavy Higgs bosons with mono-$W$ signatures in the relevant parameter planes $M_1 - \mu$ (left panel) and $M_2 - \mu$ (right panel). For higgsino-like LSPs and thus, a low $\mu$, large cross sections are found for low $M_2$ values. Therefore, the plots show a preferred coupling of heavy Higgs bosons to highly mixed higgsino-like and wino-like weakinos. Results with the highest cross sections are independent of the $M_1$ soft parameter and thus, we obtain no preferences for the bino-like component as expected.

For mono-$W$ final states in the higgsino-like LSP dataset, we obtain the highest cross sections for e.g. $M_1 = 1874.32$~GeV, $M_2 = -204.81$~GeV, $\mu = 193.26$~GeV, $\tan\beta = 24$ and $m_A = 696.49$~GeV. A representative point for mono-$Z$ final states with high cross sections in the bino-like LSP dataset would be  $M_1 = 66.98$~GeV, $M_2 = -821.75$~GeV, $\mu = 172.29$~GeV, $\tan\beta = 13.59$ and $m_A = 487.23$~GeV.

\begin{figure}[h]
\centering
\includegraphics[trim={0cm 4.5cm 0cm 0.8cm},clip,width=\textwidth]{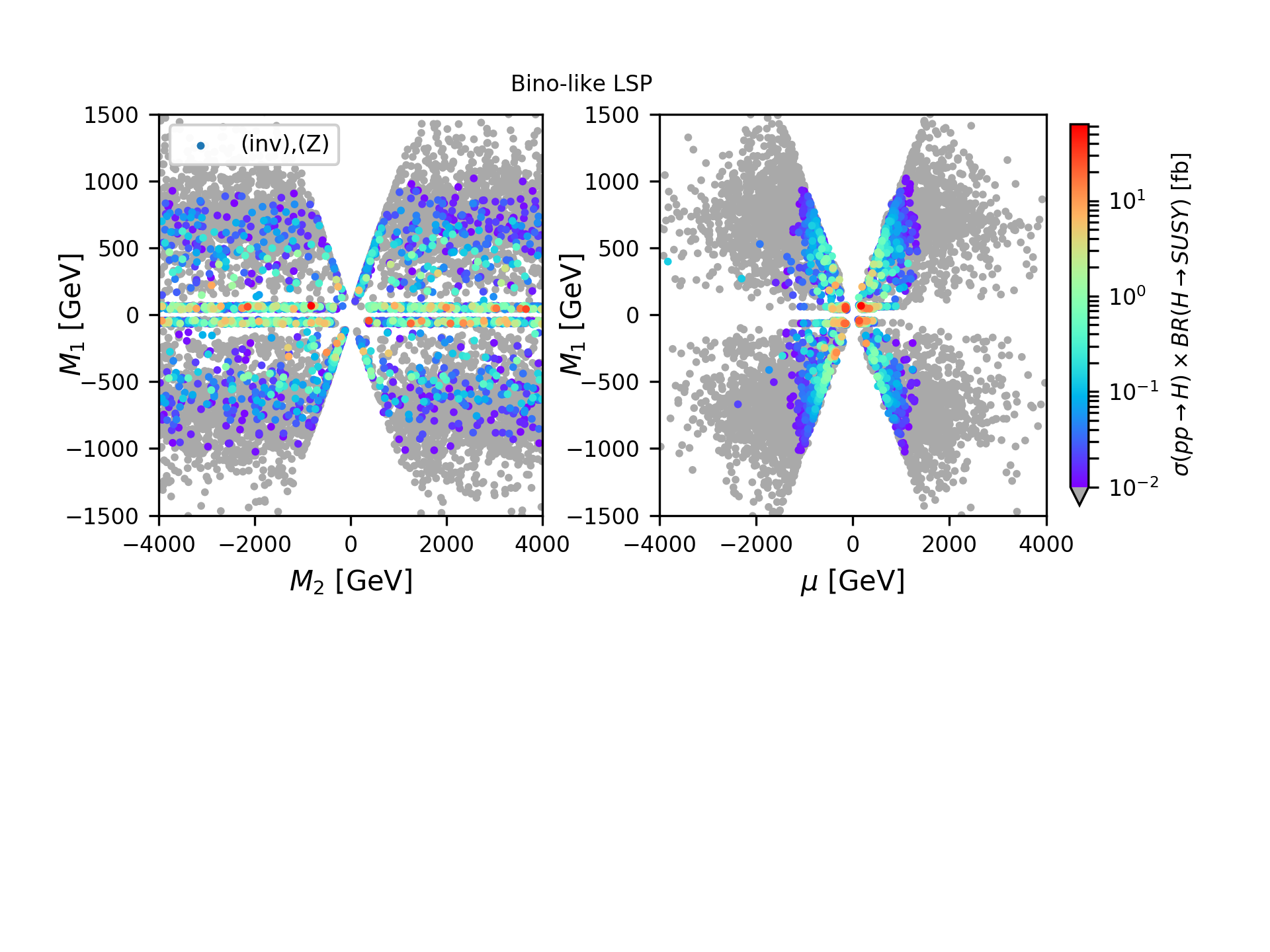}
\caption{Cross sections for heavy Higgs searches with mono-$Z$ signature in the $M_2$-$M_1$ (left) and the $\mu$-$M_1$ soft parameter plane (right) for bino-like LSPs.}\label{fig_para_3D_bino}
\end{figure}

Even though decays to wino-/higgsino-mixed weakinos are preferred, the necessity of a large gaugino-higgsino mixing for heavy Higgs couplings leads to low values of $\mu$ for bino-like LSPs (see fig.~\ref{fig_para_3D_bino}). Nevertheless, the wino-like $\tilde{\chi}^0_4$ and $\tilde{\chi}^\pm_2$ play a crucial role in the dominant signatures of heavy Higgs decays, as the decay into those particles has large branching ratios if kinematically allowed. This leads to either more complex signatures (e.g. ((Z/W),(jet,jet)),(jet,jet)) or potentially high energetic final state bosons.

\begin{figure}[h]
\centering
\includegraphics[trim={0cm 4.5cm 0cm 0.8cm},clip,width=\textwidth]{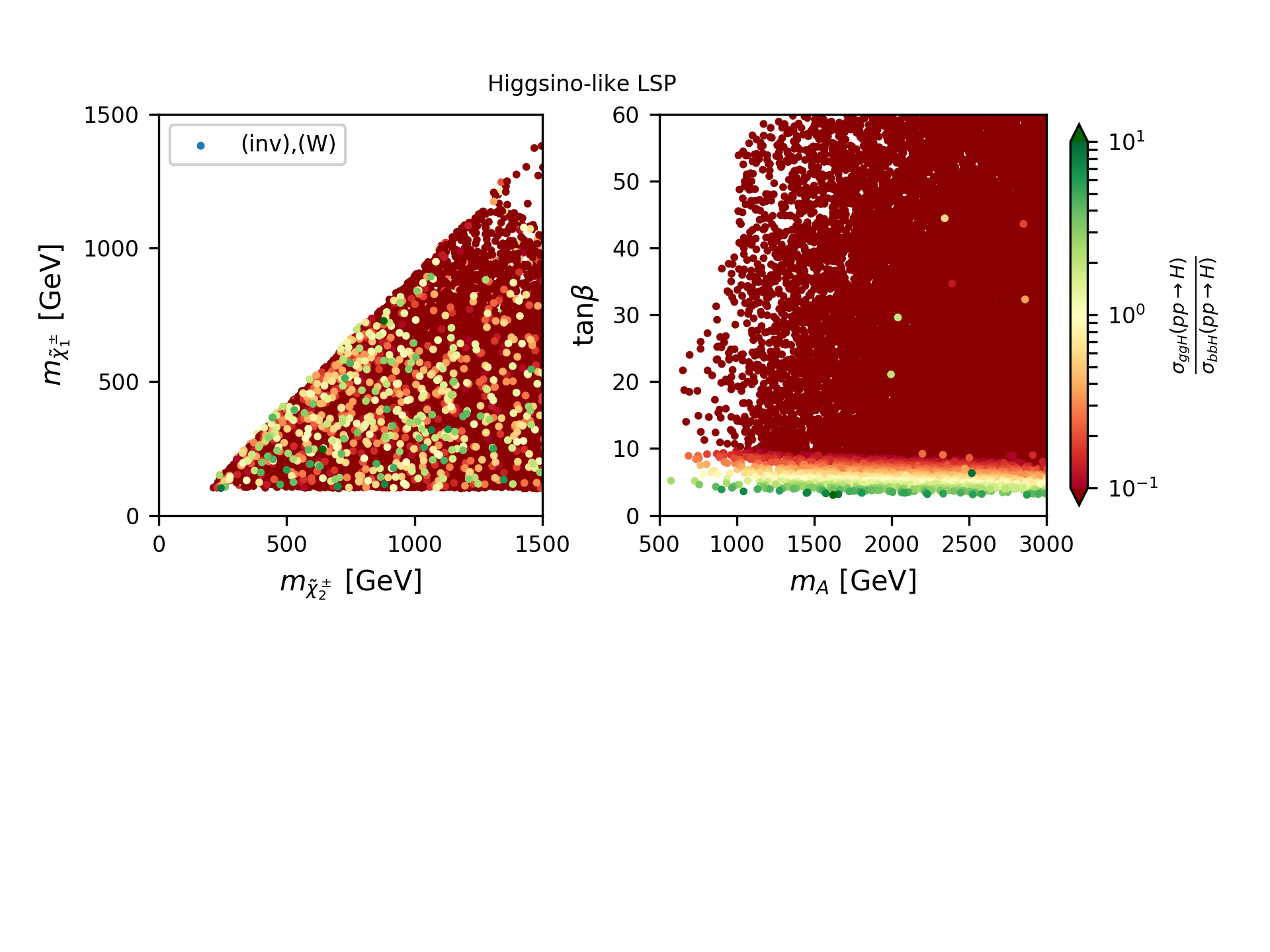}
\caption{Fraction of gluon fusion ($\sigma_{ggH}$) to bottom-quark annihilation ($\sigma_{bbH}$) cross sections for heavy Higgs production with mono-$W$ signature in the $\tilde{\chi}^\pm_2$-$\tilde{\chi}^\pm_1$ mass-plane (left) and the $m_A$-$\tan\beta$ parameter plane (right) for higgsino-like LSPs.}\label{fig_productionchannel}
\end{figure}

In particular it is interesting to analyze the production channels of the Higgs and hence possible impact on the final states. \texttt{SuSHi} takes into account gluon fusion processes ($ggH$) and bottom-quark annihilation ($bbH$) production channels to compute the heavy Higgs production cross section. It is important to note which of the production channels dominates for the mono-$X$ final states and therefore whether additional jets or $b$-quarks are to be expected in the events. Fig.~\ref{fig_productionchannel} shows the ratio of the cross sections for the two production channels from proton-proton collisions. Since the gluon fusion process is dominant for low $\tan\beta$ ($\lesssim 7$), mono-$X$ signatures from heavy Higgs decays with large cross sections are the result of $H$ being produced in bottom-quark annihilation processes. This may leave up to two potentially hard $b$-jets in the detector. Similar results are found for other mono-$X$ topologies and the bino-like LSP dataset.

The outlier parameter sets with high $\tan\beta$ and a large cross section due to gluon fusion contain light sbottom-quarks~($\tilde{b}_1, \tilde{b}_2$). Therefore the dominant gluon fusion production mechanism is the result of light sbottoms in the triangle loop diagram with its contribution being enhanced by a large $\tan\beta$ value.

\subsection*{Invisible Heavy Higgs Decays}\label{sec_inv}

For low masses of the Higgs boson, a decay to heavy SUSY particles can become kinematically forbidden. Thus, the only relevant decay products are LSP pairs, where this process results in solely missing transverse momentum in the detector. However, large cross sections of (inv),(inv) signatures are the result of asymmetric $H$ decays to nearly degenerate light neutralinos, or a pair of light charginos. The heavier states then decay softly to the LSP.

\begin{figure}[h]
\centering
\includegraphics[trim={0.5cm 4.5cm 0.5cm 0.5cm},clip,width=\textwidth]{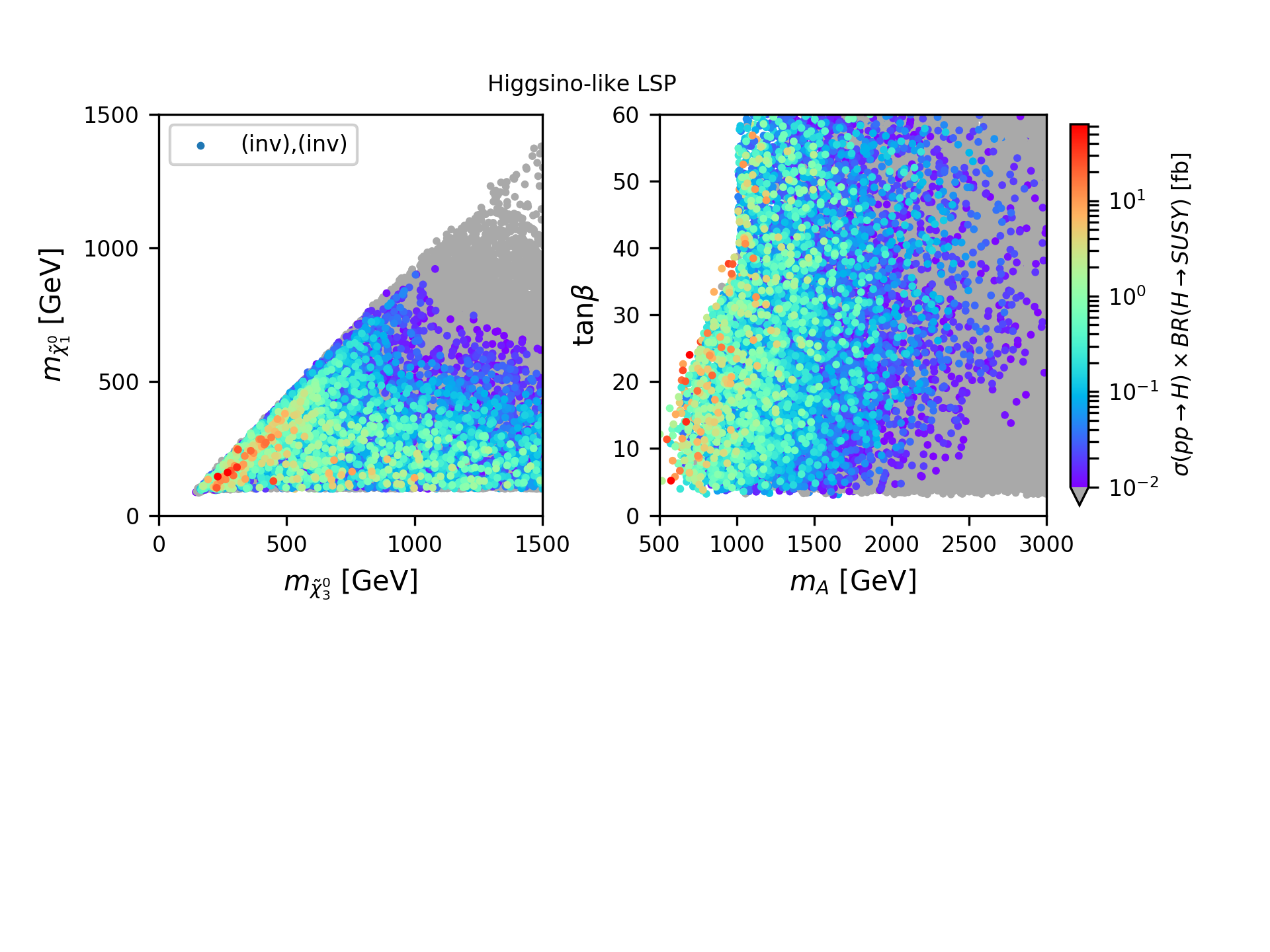}
\caption{Cross sections of invisible heavy Higgs decays in the $\tilde{\chi}^0_3-\tilde{\chi}^0_1$ mass plane (left) and the $m_A-\tan\beta$ parameter plane (right) for higgsino-like LSPs.}\label{fig_inv}
\end{figure}

In fig.~\ref{fig_inv}, we visualize the calculated cross sections for heavy Higgs productions at 14~TeV center-of-mass energy times the branching ratio leading to these signatures for the higgsino-like LSP dataset. We find similar results for the bino-like LSP dataset with potentially large cross sections for fairly light Higgs bosons. The $\sigma\times\mathcal{B}$ values reduce drastically for increasing Higgs masses, as then other decay processes become dominant. The (inv),(inv) signature is the dominant process, having the highest cross section for the majority of the bino-like LSP dataset and a significant amount of the higgsino-like LSP dataset.

In order to constrain (inv),(inv) final state, the experiments often rely on the presence of hard initial state radiation e.g. in mono-$jet$ searches. For heavy Higgs bosons produced in bottom-quark annihilation processes, the additional production of up to two potentially hard $b$-jets can thus be used in addition to the existence of soft final states and missing energy.

\subsection*{Long Cascade Decays}
\begin{figure}[h]
\centering
\includegraphics[trim={0cm 4.5cm 0cm 0.8cm},clip,width=\textwidth]{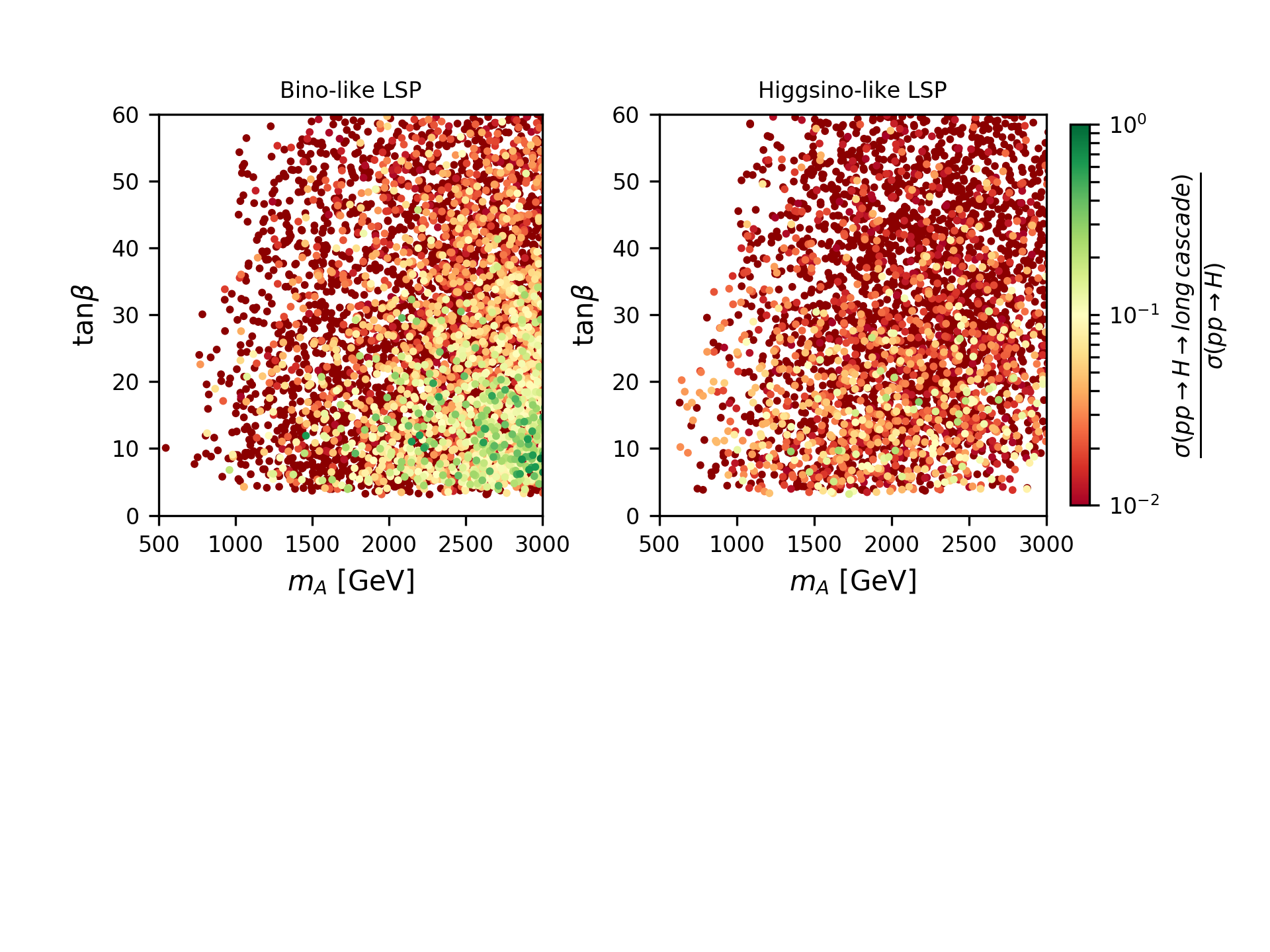}
\caption{Ratio of the cross section to long-cascade decays $\sigma(pp\rightarrow H \rightarrow$ long cascade), with more than one intermediate state, to the total cross section for resonant heavy Higgs productions $\sigma(pp\rightarrow H)$ in the $m_A$-$\tan\beta$ parameter plane for bino-like LSPs (left) and higgsino-like LSPs (right).}\label{fig_cascade}
\end{figure}

While the above analysis shows the importance of mono-$X$ topologies, it is unclear whether there are any additional complex decay chains with even larger signal cross sections in the parameter space. In order to address this question, we use the long-decay chains module of \texttt{SModelS}, to understand Higgs decays with more than one intermediate sparticle per branch. These signatures are the result of heavy Higgs decaying to heavy neutralino/chargino states. Fig.~\ref{fig_cascade} shows the ratio of the cross section to long-cascade decays~$\sigma(pp\rightarrow H \rightarrow$ long cascade) to the total cross section of heavy Higgs productions in $pp$-collisions $\sigma(pp\rightarrow H)$ in the $m_A$-$\tan\beta$ parameter plane. We find long-cascade decays are negligible in regions where we expect large Higgs production cross sections. We only obtain a non-negligible signal with long-cascade decays for the bino-like LSP dataset with large Higgs masses, where Higgs decay to the wino-like heavy neutralino and chargino has large branching ratio and thus a cascade decay is expected. Only for $m_H \approx 3$~TeV and $\tan\beta \approx 10$, the entire Higgs production cross section can lead to long cascade decays. Therefore, simple decay chains with e.g. mono-$X$ signatures in the detector are dominating the heavy Higgs boson decay to sparticles.

\subsection*{Impact of 13 TeV SUSY Searches}
The pMSSM dataset used, although is one of most complete phenomenologically viable dataset after LHC Run-1, it does not include 13 TeV LHC results. It is therefore interesting to understand the impact of 13 TeV LHC searches and any implications for Higgs decays. Studies of the pMSSM parameter set with hadronic 13 TeV LHC Run 2 ATLAS constraints from early direct SUSY searches with $3.2$~fb$^{-1}$ integrated luminosity have been performed~\cite{Barr:2016inz, Barr:2016sho}. We use the public results from these studies~\cite{pmssm:barr} to identify changes in our missing topologies. We find that hadronic analyses from 13 TeV LHC searches considered in~\cite{Barr:2016inz, Barr:2016sho} are able to exclude about $13.8~\%$ ($18.4~\%$) of the higgsino-like (bino-like) LSP dataset after applying all constraints mentioned before. The excluded SLHA files have no influence on the most frequent and most dominant missing topologies and are not able to exclude regions of the $m_A-\tan\beta$ parameter plane with definite characteristics.

\subsection*{Wino-like LSP}
\begin{figure}[h]
\includegraphics[trim={0cm 4.5cm 0cm 0.8cm}, clip, width=\textwidth]{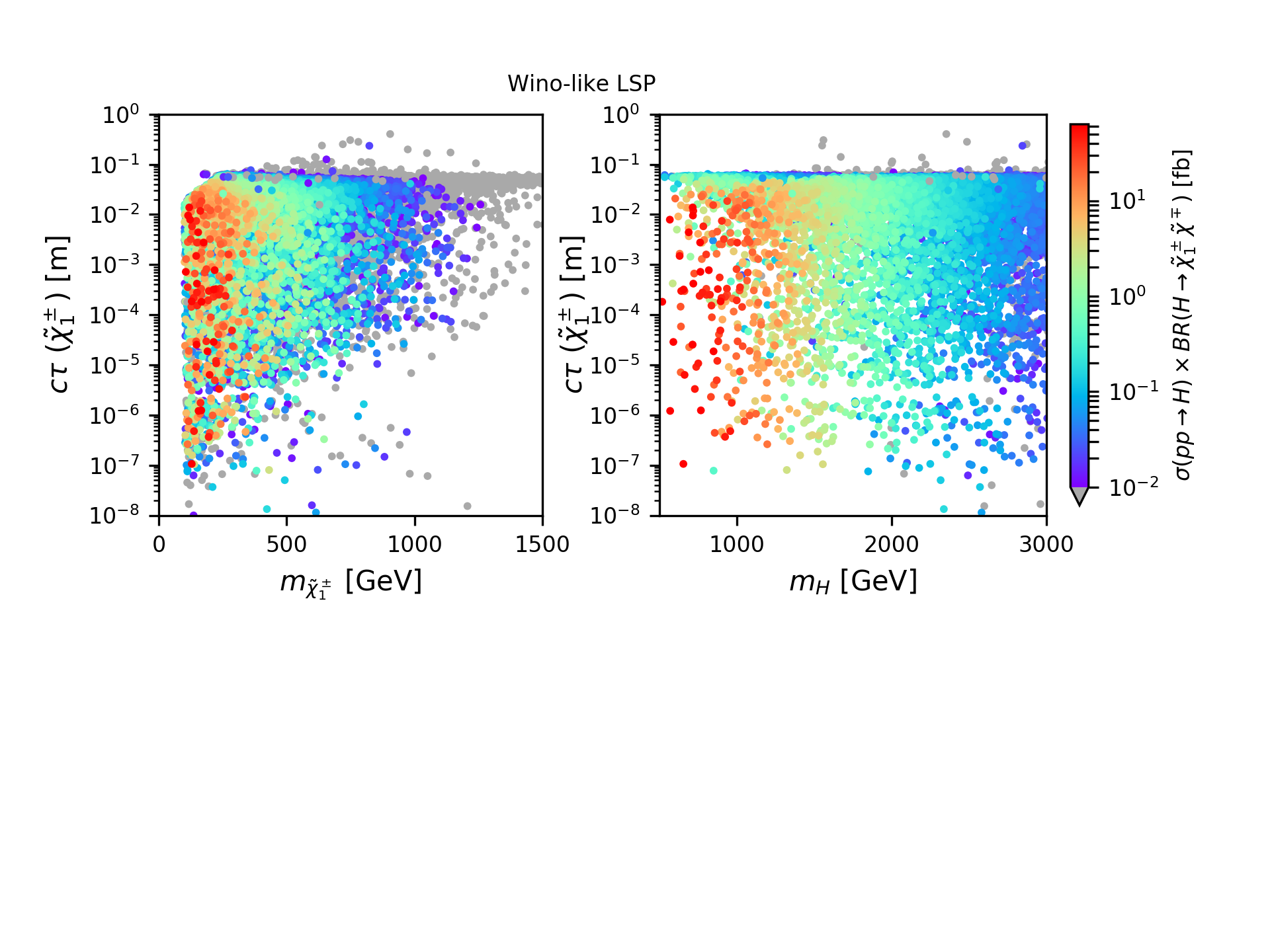}
\caption{Cross sections for heavy Higgs bosons decaying to charginos ($H\rightarrow \tilde{\chi}_1 ^\pm\tilde{\chi}_i ^\mp$, $i=1,2$) in dependence of the light chargino decay length~$c\tau(\tilde{\chi}_1^\pm)$ and its mass (left) and the heavy Higgs mass (right) for wino-like LSPs.}\label{fig_wino}
\end{figure}

The small mass difference of the LSP to the light chargino~$\tilde{\chi}_1^\pm$ leads to a long lifetime of the latter and therefore displaced vertex signatures in the detector. As the current version of \texttt{SModelS} cannot handle displaced vertices, we did not include the wino dataset in the above results. We study the cross sections in dependence of the decay length~$c\tau$ and the light chargino (heavy Higgs) mass of heavy Higgs bosons decaying to a pair of charginos including the long-lived light chargino ($H\rightarrow \tilde{\chi}_1 ^\pm\tilde{\chi}_i ^\mp$, $i=1,2$) for wino-like LSPs as seen in fig.~\ref{fig_wino} left (right). We find promising results with cross sections up to $80~\text{fb}$ for $\sqrt{s}=14$~TeV, as the branching ratio of the heavy Higgs boson decaying to charginos dominates over a decay to neutralinos and the preferred decay of heavy Higgs to a mixture of wino-/higgsino-like LSPs. A benchmark point for such high cross sections can be obtained by setting $M_1 = -1211.10$~GeV, $M_2 = -108.53$~GeV, $\mu = 388.07$~GeV, $\tan\beta = 26.50$ and $m_A = 717.71$~GeV. Thus, resonant heavy Higgs searches for wino-like LSPs with displaced vertex signatures are a potentially good candidate for future long-lived particle searches at the LHC.

\section{Conclusion}
\label{sec:conclusion}
\noindent

In this study we characterized the decays of MSSM neutral Higgs bosons to supersymmetric particles. We used a pre-collected data sample of a full 19-dimensional pMSSM parameter space. The dataset already obeys constraints from the dark matter relic density, LEP and ATLAS searches for SUSY particles, and it requires the light Higgs to be within the observed mass range. Furthermore, we have checked the parameter sets against current limits from the Higgs sector using  \texttt{HiggsBounds} and \texttt{HiggsSignals}.

We used the \texttt{SModelS} framework to analyze the phenomenologically viable parameter sets and characterize possible signatures resulting from heavy Higgs bosons decaying to supersymmetric particles. Given the large possibilities for the heavy Higgs boson to decay to supersymmetric particles and further decays of SUSY particles, it is important to use automatized approaches such as \texttt{SModelS} for efficient classification. Since \texttt{SModelS} is not able to handle resonant searches, we propose a unique method of using the framework. By providing the calculated product of the production cross section, computed with \texttt{SusHi}, and the branching ratio of the decay to SUSY particles $\sigma(pp\rightarrow H)\times\mathcal{B}(H\rightarrow SUSY)$ in the input-file, it is possible to apply \texttt{SModelS} for resonant heavy Higgs searches.

It should be noted that the methods used in this work can also be used to classify the decays of CP-odd and charged Higgs boson. The production mechanisms of CP-odd Higgs are similar to that of CP-even for the decoupling regime, however the couplings to the weakino sector depend on $\tan\beta$ and not on the Higgs mixing angle $\alpha$. The decay chains are hence expected to be different. For charged Higgs the production mechanisms are very different compared to the neutral Higgses, however once the production cross sections are computed, the methodology presented here can be readily applied.

The analysis of the heavy neutral Higgs boson for the higgsino-like and bino-like LSP datasets leads to mono-$X$ ($X$ = $h$, $W$, $Z$) signatures being the most frequently occurring as well as topologies with the largest cross sections. These signatures are the result of the asymmetric decay of $H$ to a pair of neutralinos or charginos, where the lighter state decays softly to the LSP ($\Delta m < 5$~GeV) and the heavier state decays to the LSP and $X$. The calculated production cross section resulting in those signatures revealed values up to $10$~fb for a center-of-mass energy of $\sqrt{s}=14$~TeV. 

Furthermore, a detailed study of their underlying parameter sets showed large cross sections for highly mixed weakino states, preferring mixed higgsino-/wino-like LSPs. Mono-$X$ signatures will have additional $b$-jets in the detector as a result of bottom-quark annihilation being the dominant heavy Higgs production process for about $98~\%$ of the analyzed pMSSM parameter space. Since we found cross sections with long-cascade decay signatures being negligible, the decay of heavy Higgs bosons to sparticles will dominantly result in simple topologies.

Studies performed in Ref.~\cite{Barr:2016inz} and \cite{Barr:2016sho} used early results from LHC Run 2 with $3.2$~fb$^{-1}$ integrated luminosity to additionally constrain the pMSSM dataset. We applied this study to show the possible exclusion of $13.8~\%$ ($18.4~\%$) of the remaining higgsino-like (bino-like) LSP dataset. However, these constrains are not able to exclude a certain region of the pMSSM parameter space and have no influence on the most frequent and dominant missing topologies of heavy Higgs decays.

We also argue that solely by considering direct Higgs searches, where the heavy Higgs decays to SM particles, is insufficient to probe the full pMSSM parameter space. Therefore, resonant heavy Higgs searches with a decay to SUSY particles are crucial for future analyses.

As we obtain a preferred decay of the heavy Higgs boson to highly mixed higgsino-/wino-like LSPs, we expect large cross sections in the wino-like LSP dataset. However, the small mass difference of the LSP to the light chargino~$\tilde{\chi}_1^\pm$ results in displaced vertex signatures that cannot be  handled by \texttt{SModelS} yet. A brief analysis showed expected $\sigma(pp\rightarrow H)\times\mathcal{B}(H\rightarrow \tilde{\chi}^\pm\tilde{\chi}^\mp)$ values for decays to a pair of charginos of up to $80$~fb for a center-of-mass energy of $\sqrt{s}=14$~TeV. 

Even though the heavy Higgs boson production process in proton-proton collisions as well as its decay to supersymmetric particles is highly dependent on the underlying parameters of the 19-dimensional pMSSM parameter space, it mainly results in simple signatures, as e.g. mono-$X$ ($X$ = $h$, $W$, $Z$). We found potentially large cross sections for those topologies being reachable for the future LHC update. The high-luminosity LHC, with a total of $3~\text{ab}^{-1}$ integrated luminosity will therefore be able to probe the pMSSM parameter space by searching for heavy Higgs bosons decaying to sparticles. 

Given the results of this study, it will be interesting to reinterpret the existing mono-$X$ searches and demonstrate the existing coverage for heavy Higgs decay. This will be covered in a future work. For the limited number of points resulting in signatures other than mono-$X$ final states, it is interesting to ask whether the signal shape is different as compared to SUSY production via SM mediators. Such differences in the signal shapes can then be used to search for resonance production decaying to supersymmetric particles. This question will be addressed in an upcoming study.
 
\section*{Acknowledgements}
LL is partially supported by the Austrian FWF, project P26896-N27. SK is supported by the `New Frontiers' program of the Austrian Academy of Sciences and by FWF project number V592-N27. We thank the HEPHY CMS group members for numerous discussions. We furthermore thank members of the \texttt{SModelS} collaboration for discussions. We in particular thank U. Laa for useful discussions and comments on the manuscript. We also thank F. Deppisch for careful reading of this manuscript and his invaluable comments.
\bibliographystyle{JHEP}
\bibliography{bibliography}

\providecommand{\href}[2]{#2}\begingroup\raggedright\begin{thebibliography}{100}

\bibitem{Aad:2012tfa}
{\bf ATLAS} Collaboration, G.~Aad et~al., {\it {Observation of a new particle
  in the search for the Standard Model Higgs boson with the ATLAS detector at
  the LHC}},  {\em Phys. Lett.} {\bf B716} (2012) 1--29,
  [\href{http://arxiv.org/abs/1207.7214}{{\tt arXiv:1207.7214}}].

\bibitem{Chatrchyan:2012xdj}
{\bf CMS} Collaboration, S.~Chatrchyan et~al., {\it {Observation of a new boson
  at a mass of 125 GeV with the CMS experiment at the LHC}},  {\em Phys. Lett.}
  {\bf B716} (2012) 30--61, [\href{http://arxiv.org/abs/1207.7235}{{\tt
  arXiv:1207.7235}}].

\bibitem{Branco:2011iw}
G.~C. Branco, P.~M. Ferreira, L.~Lavoura, M.~N. Rebelo, M.~Sher, and J.~P.
  Silva, {\it {Theory and phenomenology of two-Higgs-doublet models}},  {\em
  Phys. Rept.} {\bf 516} (2012) 1--102,
  [\href{http://arxiv.org/abs/1106.0034}{{\tt arXiv:1106.0034}}].

\bibitem{LopezHonorez:2006gr}
L.~Lopez~Honorez, E.~Nezri, J.~F. Oliver, and M.~H.~G. Tytgat, {\it {The Inert
  Doublet Model: An Archetype for Dark Matter}},  {\em JCAP} {\bf 0702} (2007)
  028, [\href{http://arxiv.org/abs/hep-ph/0612275}{{\tt hep-ph/0612275}}].

\bibitem{Drees:873465}
M.~Drees, R.~Godbole, and P.~Roy, {\it {Theory and phenomenology of Sparticles:
  an account of four-dimensional N=1 supersymmetry in high-energy physics}},
  {\em World Scientific} (2004).

\bibitem{Gunion:1989we}
J.~F. Gunion, H.~E. Haber, G.~L. Kane, and S.~Dawson, {\it {The Higgs Hunter's
  Guide}},  {\em Front. Phys.} {\bf 80} (2000) 1--404.

\bibitem{Djouadi:2005gj}
A.~Djouadi, {\it {The Anatomy of electro-weak symmetry breaking. II. The Higgs
  bosons in the minimal supersymmetric model}},  {\em Phys. Rept.} {\bf 459}
  (2008) 1--241, [\href{http://arxiv.org/abs/hep-ph/0503173}{{\tt
  hep-ph/0503173}}].

\bibitem{Martin:1997ns}
S.~P. Martin, {\it {A Supersymmetry primer}},
  \href{http://arxiv.org/abs/hep-ph/9709356}{{\tt hep-ph/9709356}}. [Adv. Ser.
  Direct. High Energy Phys.18,1(1998)].

\bibitem{Barman:2016kgt}
R.~K. Barman, B.~Bhattacherjee, A.~Chakraborty, and A.~Choudhury, {\it {Study
  of MSSM heavy Higgs bosons decaying into charginos and neutralinos}},  {\em
  Phys. Rev.} {\bf D94} (2016), no.~7 075013,
  [\href{http://arxiv.org/abs/1607.00676}{{\tt arXiv:1607.00676}}].

\bibitem{Arbey:2013jla}
A.~Arbey, M.~Battaglia, and F.~Mahmoudi, {\it {Supersymmetric Heavy Higgs
  Bosons at the LHC}},  {\em Phys. Rev.} {\bf D88} (2013), no.~1 015007,
  [\href{http://arxiv.org/abs/1303.7450}{{\tt arXiv:1303.7450}}].

\bibitem{Belanger:2015vwa}
G.~Belanger, D.~Ghosh, R.~Godbole, and S.~Kulkarni, {\it {Light stop in the
  MSSM after LHC Run 1}},  {\em JHEP} {\bf 09} (2015) 214,
  [\href{http://arxiv.org/abs/1506.00665}{{\tt arXiv:1506.00665}}].

\bibitem{Djouadi:1996mj}
A.~Djouadi, P.~Janot, J.~Kalinowski, and P.~M. Zerwas, {\it {SUSY decays of
  Higgs particles}},  {\em Phys. Lett.} {\bf B376} (1996) 220--226,
  [\href{http://arxiv.org/abs/hep-ph/9603368}{{\tt hep-ph/9603368}}].

\bibitem{Bisset:2007mk}
M.~Bisset, J.~Li, and N.~Kersting, {\it {How to Detect `Decoupled' Heavy
  Supersymmetric Higgs Bosons}},  \href{http://arxiv.org/abs/0709.1031}{{\tt
  arXiv:0709.1031}}.

\bibitem{Arhrib:2011rp}
A.~Arhrib, R.~Benbrik, M.~Chabab, and C.-H. Chen, {\it {Pair production of
  neutralinos and charginos at the LHC: the role of Higgs bosons exchange}},
  {\em Phys. Rev.} {\bf D84} (2011) 115012,
  [\href{http://arxiv.org/abs/1109.0518}{{\tt arXiv:1109.0518}}].

\bibitem{Belanger:2000tg}
G.~Belanger, F.~Boudjema, F.~Donato, R.~Godbole, and S.~Rosier-Lees, {\it {SUSY
  Higgs at the LHC: Effects of light charginos and neutralinos}},  {\em Nucl.
  Phys.} {\bf B581} (2000) 3--33,
  [\href{http://arxiv.org/abs/hep-ph/0002039}{{\tt hep-ph/0002039}}].

\bibitem{Bisset:2000ud}
M.~Bisset, M.~Guchait, and S.~Moretti, {\it {Signatures of MSSM charged Higgs
  bosons via chargino neutralino decay channels at the LHC}},  {\em Eur. Phys.
  J.} {\bf C19} (2001) 143--154,
  [\href{http://arxiv.org/abs/hep-ph/0010253}{{\tt hep-ph/0010253}}].

\bibitem{Medina:2017bke}
A.~D. Medina and M.~A. Schmidt, {\it {Enlarging Regions of the MSSM Parameter
  Space for Large $\tan\beta$ via SUSY Decays of the Heavy Higgs Bosons}},
  {\em JHEP} {\bf 08} (2017) 095, [\href{http://arxiv.org/abs/1706.04994}{{\tt
  arXiv:1706.04994}}].

\bibitem{Barman:2016jov}
R.~K. Barman, B.~Bhattacherjee, A.~Choudhury, D.~Chowdhury, J.~Lahiri, and
  S.~Ray, {\it {Current status of MSSM Higgs sector with LHC 13 TeV data}},
  \href{http://arxiv.org/abs/1608.02573}{{\tt arXiv:1608.02573}}.

\bibitem{Bisset:2007mi}
M.~Bisset, J.~Li, N.~Kersting, R.~Lu, F.~Moortgat, and S.~Moretti, {\it
  {Four-lepton LHC events from MSSM Higgs boson decays into neutralino and
  chargino pairs}},  {\em JHEP} {\bf 08} (2009) 037,
  [\href{http://arxiv.org/abs/0709.1029}{{\tt arXiv:0709.1029}}].

\bibitem{Gunion:1988yc}
J.~F. Gunion and H.~E. Haber, {\it {Higgs Bosons in Supersymmetric Models. 3.
  Decays Into Neutralinos and Charginos}},  {\em Nucl. Phys.} {\bf B307} (1988)
  445. [Erratum: Nucl. Phys.B402,569(1993)].

\bibitem{Li:2013nma}
T.~Li, {\it {Decoupling MSSM Higgs Sector and Heavy Higgs Decay}},  {\em Phys.
  Lett.} {\bf B728} (2014) 77--84, [\href{http://arxiv.org/abs/1309.6713}{{\tt
  arXiv:1309.6713}}].

\bibitem{Ananthanarayan:2015fwa}
B.~Ananthanarayan, J.~Lahiri, and P.~N. Pandita, {\it {Invisible decays of the
  heavier Higgs boson in the minimal supersymmetric standard model}},  {\em
  Phys. Rev.} {\bf D91} (2015) 115025,
  [\href{http://arxiv.org/abs/1507.01747}{{\tt arXiv:1507.01747}}].

\bibitem{Djouadi:2015jea}
A.~Djouadi, L.~Maiani, A.~Polosa, J.~Quevillon, and V.~Riquer, {\it {Fully
  covering the MSSM Higgs sector at the LHC}},  {\em JHEP} {\bf 06} (2015) 168,
  [\href{http://arxiv.org/abs/1502.05653}{{\tt arXiv:1502.05653}}].

\bibitem{Zhang:2002fu}
R.-Y. Zhang, W.-G. Ma, L.-H. Wan, and Y.~Jiang, {\it {Supersymmetric
  electroweak corrections to the Higgs boson decays into chargino or neutralino
  pair}},  {\em Phys. Rev.} {\bf D65} (2002) 075018,
  [\href{http://arxiv.org/abs/hep-ph/0201132}{{\tt hep-ph/0201132}}].

\bibitem{Ibrahim:2008rq}
T.~Ibrahim, {\it {Neutralino decay of MSSM neutral Higgs bosons}},  {\em Phys.
  Rev.} {\bf D77} (2008) 065028, [\href{http://arxiv.org/abs/0803.4134}{{\tt
  arXiv:0803.4134}}].

\bibitem{Ambrogi:2017neo}
F.~Ambrogi et~al., {\it {SModelS v1.1 user manual}},
  \href{http://arxiv.org/abs/1701.06586}{{\tt arXiv:1701.06586}}.

\bibitem{Kraml:2013mwa}
S.~Kraml et~al., {\it {SModelS: a tool for interpreting simplified-model
  results from the LHC and its application to supersymmetry}},  {\em Eur. Phys.
  J.} {\bf C74} (2014) 2868, [\href{http://arxiv.org/abs/1312.4175}{{\tt
  arXiv:1312.4175}}].

\bibitem{SModelS:wiki}
{http://smodels.hephy.at}.

\bibitem{Aad:2015baa}
{\bf ATLAS} Collaboration, G.~Aad et~al., {\it {Summary of the ATLAS
  experiment?s sensitivity to supersymmetry after LHC Run 1 ? interpreted in
  the phenomenological MSSM}},  {\em JHEP} {\bf 10} (2015) 134,
  [\href{http://arxiv.org/abs/1508.06608}{{\tt arXiv:1508.06608}}].

\bibitem{Berger:2008cq}
C.~F. Berger, J.~S. Gainer, J.~L. Hewett, and T.~G. Rizzo, {\it {Supersymmetry
  Without Prejudice}},  {\em JHEP} {\bf 02} (2009) 023,
  [\href{http://arxiv.org/abs/0812.0980}{{\tt arXiv:0812.0980}}].

\bibitem{CahillRowley:2012cb}
M.~W. Cahill-Rowley, J.~L. Hewett, S.~Hoeche, A.~Ismail, and T.~G. Rizzo, {\it
  {The New Look pMSSM with Neutralino and Gravitino LSPs}},  {\em Eur. Phys.
  J.} {\bf C72} (2012) 2156, [\href{http://arxiv.org/abs/1206.4321}{{\tt
  arXiv:1206.4321}}].

\bibitem{CahillRowley:2012kx}
M.~W. Cahill-Rowley, J.~L. Hewett, A.~Ismail, and T.~G. Rizzo, {\it {More
  energy, more searches, but the phenomenological MSSM lives on}},  {\em Phys.
  Rev.} {\bf D88} (2013), no.~3 035002,
  [\href{http://arxiv.org/abs/1211.1981}{{\tt arXiv:1211.1981}}].

\bibitem{Cahill-Rowley:2014twa}
M.~Cahill-Rowley, J.~L. Hewett, A.~Ismail, and T.~G. Rizzo, {\it {Lessons and
  prospects from the pMSSM after LHC Run I}},  {\em Phys. Rev.} {\bf D91}
  (2015), no.~5 055002, [\href{http://arxiv.org/abs/1407.4130}{{\tt
  arXiv:1407.4130}}].

\bibitem{Bechtle:2013xfa}
P.~Bechtle et~al., {\it {$HiggsSignals$: Confronting arbitrary Higgs sectors
  with measurements at the Tevatron and the LHC}},  {\em Eur. Phys. J.} {\bf
  C74} (2014), no.~2 2711, [\href{http://arxiv.org/abs/1305.1933}{{\tt
  arXiv:1305.1933}}].

\bibitem{Stal:2013hwa}
O.~Stål and T.~Stefaniak, {\it {Constraining extended Higgs sectors with
  HiggsSignals}},  {\em PoS} {\bf EPS-HEP2013} (2013) 314,
  [\href{http://arxiv.org/abs/1310.4039}{{\tt arXiv:1310.4039}}].

\bibitem{Bechtle:2013gu}
P.~Bechtle et~al., {\it {Recent Developments in HiggsBounds and a Preview of
  HiggsSignals}},  {\em PoS} {\bf CHARGED2012} (2012) 024,
  [\href{http://arxiv.org/abs/1301.2345}{{\tt arXiv:1301.2345}}].

\bibitem{arXiv:0811.4169}
P.~Bechtle, O.~Brein, S.~Heinemeyer, G.~Weiglein, and K.~E. Williams, {\it
  {HiggsBounds: Confronting Arbitrary Higgs Sectors with Exclusion Bounds from
  LEP and the Tevatron}},  {\em Comput. Phys. Commun.} {\bf 181} (2010)
  138--167, [\href{http://arxiv.org/abs/0811.4169}{{\tt arXiv:0811.4169}}].

\bibitem{arXiv:1102.1898}
P.~Bechtle, O.~Brein, S.~Heinemeyer, G.~Weiglein, and K.~E. Williams, {\it
  {HiggsBounds 2.0.0: Confronting Neutral and Charged Higgs Sector Predictions
  with Exclusion Bounds from LEP and the Tevatron}},  {\em Comput. Phys.
  Commun.} {\bf 182} (2011) 2605--2631,
  [\href{http://arxiv.org/abs/1102.1898}{{\tt arXiv:1102.1898}}].

\bibitem{arXiv:1301.2345}
P.~Bechtle et~al., {\it {Recent Developments in HiggsBounds and a Preview of
  HiggsSignals}},  {\em PoS} {\bf CHARGED2012} (2012) 024,
  [\href{http://arxiv.org/abs/1301.2345}{{\tt arXiv:1301.2345}}].

\bibitem{arXiv:1311.0055}
P.~Bechtle et~al., {\it {HiggsBounds-4: Improved Tests of Extended Higgs
  Sectors against Exclusion Bounds from LEP, the Tevatron and the LHC}},  {\em
  Eur. Phys. J.} {\bf C74} (2014) 2693,
  [\href{http://arxiv.org/abs/1311.0055}{{\tt arXiv:1311.0055}}].

\bibitem{arXiv:1507.06706}
P.~Bechtle, S.~Heinemeyer, O.~Stal, T.~Stefaniak, and G.~Weiglein, {\it
  {Applying Exclusion Likelihoods from LHC Searches to Extended Higgs
  Sectors}},  \href{http://arxiv.org/abs/1507.06706}{{\tt arXiv:1507.06706}}.

\bibitem{arXiv:1207.7214}
{\bf ATLAS} Collaboration, {\it {Observation of a new particle in the search
  for the Standard Model Higgs boson with the ATLAS detector at the LHC}},
  {\em Phys. Lett.} {\bf B716} (2012) 1--29,
  [\href{http://arxiv.org/abs/1207.7214}{{\tt arXiv:1207.7214}}].

\bibitem{arXiv:1011.1931}
{\bf D0} Collaboration, V.~M. Abazov et~al., {\it {Search for neutral Higgs
  bosons in the multi-b-jet topology in 5.2fb-1 of ppbar collisions at
  sqrt(s)=1.96 TeV}},  {\em Phys. Lett.} {\bf B698} (2011) 97--104,
  [\href{http://arxiv.org/abs/1011.1931}{{\tt arXiv:1011.1931}}].

\bibitem{hep-ex/0206022}
{\bf OPAL} Collaboration, G.~Abbiendi et~al., {\it {Decay-mode independent
  searches for new scalar bosons with the OPAL detector at LEP}},  {\em Eur.
  Phys. J.} {\bf C27} (2003) 311--329,
  [\href{http://arxiv.org/abs/hep-ex/0206022}{{\tt hep-ex/0206022}}].

\bibitem{arxiv:1202.1488}
{\bf CMS} Collaboration, S.~Chatrchyan, {\it {Combined results of searches for
  the standard model Higgs boson in pp collisions at sqrt(s) = 7 TeV}},  {\em
  Phys. Lett.} {\bf B710} (2012) 26--48,
  [\href{http://arxiv.org/abs/1202.1488}{{\tt arXiv:1202.1488}}].

\bibitem{arXiv:1008.3564}
{\bf D0} Collaboration, V.~M. Abazov et~al., {\it {Search for $ZH \rightarrow
  \ell^+\ell^-b\bar{b}$ production in $4.2$~fb$^{-1}$ of $p\bar{p}$ collisions
  at $\sqrt{s}=1.96$ TeV}},  {\em Phys. Rev. Lett.} {\bf 105} (2010) 251801,
  [\href{http://arxiv.org/abs/1008.3564}{{\tt arXiv:1008.3564}}].

\bibitem{arXiv:1106.4782}
{\bf CDF} Collaboration, T.~Aaltonen et~al., {\it {Search for Higgs Bosons
  Produced in Association with b- Quarks}},  {\em Phys. Rev.} {\bf D85} (2012)
  032005, [\href{http://arxiv.org/abs/1106.4782}{{\tt arXiv:1106.4782}}].

\bibitem{arXiv:1108.3331}
{\bf CDF} Collaboration, D.~Benjamin et~al., {\it {Combined CDF and D0 upper
  limits on gg->H W+W- and constraints on the Higgs boson mass in
  fourth-generation fermion models with up to 8.2 fb-1 of data}},
  \href{http://arxiv.org/abs/1108.3331}{{\tt arXiv:1108.3331}}.

\bibitem{arxiv:1202.1997}
{\bf CMS} Collaboration, S.~Chatrchyan, {\it {Search for the standard model
  Higgs boson in the decay channel H to ZZ to 4 leptons in pp collisions at
  sqrt(s) = 7 TeV}},  {\em Phys. Rev. Lett.} {\bf 108} (2012) 111804,
  [\href{http://arxiv.org/abs/1202.1997}{{\tt arXiv:1202.1997}}].

\bibitem{arXiv:0707.0373}
{\bf OPAL} Collaboration, G.~Abbiendi et~al., {\it {Search for invisibly
  decaying Higgs bosons in $e^+e^- \to Z^0 h^0$ production at $\sqrt{s}$=183 -
  209 GeV}},  {\em Phys. Lett.} {\bf B682} (2010) 381--390,
  [\href{http://arxiv.org/abs/0707.0373}{{\tt arXiv:0707.0373}}].

\bibitem{arXiv:1506.02301}
{\bf CMS} Collaboration, {\it {Search for diphoton resonances in the mass range
  from 150 to 850 GeV in pp collisions at sqrt(s) = 8 TeV}},  {\em Phys. Lett.}
  {\bf B750} (2015) 494, [\href{http://arxiv.org/abs/1506.02301}{{\tt
  arXiv:1506.02301}}].

\bibitem{arXiv:0806.0611}
{\bf D0} Collaboration, V.~M. Abazov et~al., {\it {Search for a scalar or
  vector particle decaying into Zgamma in p anti-p collisions at s**(1/2) =
  1.96-TeV}},  {\em Phys. Lett.} {\bf B671} (2009) 349--355,
  [\href{http://arxiv.org/abs/0806.0611}{{\tt arXiv:0806.0611}}].

\bibitem{hep-ex/0111010}
{\bf OPAL} Collaboration, G.~Abbiendi et~al., {\it {Search for Yukawa
  production of a light neutral Higgs boson at LEP}},  {\em Eur. Phys. J.} {\bf
  C23} (2002) 397--407, [\href{http://arxiv.org/abs/hep-ex/0111010}{{\tt
  hep-ex/0111010}}].

\bibitem{arXiv:1402.3051}
{\bf ATLAS} Collaboration, {\it {Search for Higgs boson decays to a photon and
  a Z boson in pp collisions at sqrt(s)=7 and 8 TeV with the ATLAS detector}},
  {\em Phys. Lett.} {\bf B732} (2014) 8--27,
  [\href{http://arxiv.org/abs/1402.3051}{{\tt arXiv:1402.3051}}].

\bibitem{arXiv:1107.1268}
{\bf D0} Collaboration, V.~M. Abazov et~al., {\it {Search for associated Higgs
  boson production using like charge dilepton events in ppbar collisions at
  sqrt{s} = 1.96 TeV}},  {\em Phys. Rev.} {\bf D84} (2011) 092002,
  [\href{http://arxiv.org/abs/1107.1268}{{\tt arXiv:1107.1268}}].

\bibitem{hep-ex/0501033}
{\bf L3} Collaboration, P.~Achard et~al., {\it {Search for an
  invisibly-decaying Higgs boson at LEP}},  {\em Phys. Lett.} {\bf B609} (2005)
  35--48, [\href{http://arxiv.org/abs/hep-ex/0501033}{{\tt hep-ex/0501033}}].

\bibitem{arXiv:1509.04670}
{\bf ATLAS} Collaboration, {\it {Searches for Higgs boson pair production in
  the $hh\to bb\tau\tau, \gamma\gamma WW*, \gamma\gamma bb, bbbb$ channels with
  the ATLAS detector}},  \href{http://arxiv.org/abs/1509.04670}{{\tt
  arXiv:1509.04670}}.

\bibitem{arXiv:1012.0874}
{\bf D0} Collaboration, V.~M. Abazov et~al., {\it {Search for $WH$ associated
  production in 5.3 fb$^{-1}$ of $p\bar{p}$ collisions at the Fermilab
  Tevatron}},  {\em Phys. Lett.} {\bf B698} (2011) 6--13,
  [\href{http://arxiv.org/abs/1012.0874}{{\tt arXiv:1012.0874}}].

\bibitem{arXiv:1402.3244}
{\bf ATLAS} Collaboration, {\it {Search for Invisible Decays of a Higgs Boson
  Produced in Association with a Z Boson in ATLAS}},
  \href{http://arxiv.org/abs/1402.3244}{{\tt arXiv:1402.3244}}.

\bibitem{arxiv:1112.2577}
{\bf ATLAS} Collaboration, G.~Aad, {\it {Search for the Higgs boson in the
  H->WW(*)->lvlv decay channel in pp collisions at sqrt{s} = 7 TeV with the
  ATLAS detector}},  {\em Phys. Rev. Lett.} {\bf 108} (2012) 111802,
  [\href{http://arxiv.org/abs/1112.2577}{{\tt arXiv:1112.2577}}].

\bibitem{arXiv:0908.1811}
{\bf D0} Collaboration, V.~M. Abazov et~al., {\it {Search for charged Higgs
  bosons in top quark decays}},  {\em Phys. Lett.} {\bf B682} (2009) 278--286,
  [\href{http://arxiv.org/abs/0908.1811}{{\tt arXiv:0908.1811}}].

\bibitem{arXiv:1307.5515}
{\bf CMS} Collaboration, {\it {Search for a Higgs boson decaying into a Z and a
  photon in pp collisions at sqrt(s) = 7 and 8 TeV}},  {\em Phys. Lett.} {\bf
  B726} (2013) 587, [\href{http://arxiv.org/abs/1307.5515}{{\tt
  arXiv:1307.5515}}].

\bibitem{Abazov:2010ct}
{\bf D0} Collaboration, V.~M. Abazov et~al., {\it {Search for Higgs boson
  production in dilepton and missing energy final states with 5.4 $fb^{-1}$ of
  $p\bar{p}$ collisions at ${\sqrt s =1.96}$ TeV}},  {\em Phys. Rev. Lett.}
  {\bf 104} (2010) 061804, [\href{http://arxiv.org/abs/1001.4481}{{\tt
  arXiv:1001.4481}}].

\bibitem{arxiv:1202.3478}
{\bf CMS} Collaboration, S.~Chatrchyan, {\it {Search for the standard model
  Higgs boson in the H to ZZ to 2l 2nu channel in pp collisions at sqrt(s) = 7
  TeV}},  {\em JHEP} {\bf 03} (2012) 040,
  [\href{http://arxiv.org/abs/1202.3478}{{\tt arXiv:1202.3478}}].

\bibitem{hep-ex/0410017}
{\bf DELPHI} Collaboration, J.~Abdallah et~al., {\it {Searches for neutral
  Higgs bosons in extended models}},  {\em Eur. Phys. J.} {\bf C38} (2004)
  1--28, [\href{http://arxiv.org/abs/hep-ex/0410017}{{\tt hep-ex/0410017}}].

\bibitem{arXiv:1404.1344}
{\bf CMS} Collaboration, {\it {Search for invisible decays of Higgs bosons in
  the vector boson fusion and associated ZH production modes}},  {\em Eur.
  Phys. J.} {\bf C74} (2014) 2980, [\href{http://arxiv.org/abs/1404.1344}{{\tt
  arXiv:1404.1344}}].

\bibitem{arXiv:1202.1416}
{\bf CMS} Collaboration, S.~Chatrchyan, {\it {Search for a Higgs boson in the
  decay channel H to ZZ(*) to q qbar l-l+ in pp collisions at sqrt(s) = 7
  TeV}},  {\em JHEP} {\bf 04} (2012) 036,
  [\href{http://arxiv.org/abs/1202.1416}{{\tt arXiv:1202.1416}}].

\bibitem{arXiv:1406.7663}
{\bf ATLAS} Collaboration, {\it {Search for the Standard Model Higgs boson
  decay to $\mu^{+}\mu^{-}$ with the ATLAS detector}},  {\em Physics Letters}
  {\bf B738} (2014) 68--86, [\href{http://arxiv.org/abs/1406.7663}{{\tt
  arXiv:1406.7663}}].

\bibitem{hep-ex/0107034}
{\bf LEP Higgs Working Group for Higgs boson searches} Collaboration, {\it
  {Flavor independent search for hadronically decaying neutral Higgs bosons at
  LEP}},  \href{http://arxiv.org/abs/hep-ex/0107034}{{\tt hep-ex/0107034}}.

\bibitem{arXiv:1204.2760}
{\bf ATLAS} Collaboration, G.~Aad, {\it {Search for charged Higgs bosons
  decaying via H+ -> tau nu in top quark pair events using pp collision data at
  sqrt(s) = 7 TeV with the ATLAS detector}},  {\em JHEP} {\bf 06} (2012) 039,
  [\href{http://arxiv.org/abs/1204.2760}{{\tt arXiv:1204.2760}}].

\bibitem{arXiv:1504.00936}
{\bf CMS} Collaboration, {\it {Search for a Higgs boson in the mass range from
  145 to 1000 GeV decaying to a pair of W or Z bosons}},
  \href{http://arxiv.org/abs/1504.00936}{{\tt arXiv:1504.00936}}.

\bibitem{arXiv:1108.5064}
{\bf ATLAS} Collaboration, G.~Aad, {\it {Search for a heavy Standard Model
  Higgs boson in the channel H->ZZ->llqq using the ATLAS detector}},  {\em
  Phys. Lett.} {\bf B707} (2012) 27--45,
  [\href{http://arxiv.org/abs/1108.5064}{{\tt arXiv:1108.5064}}].

\bibitem{arXiv:0906.1014}
{\bf CDF} Collaboration, T.~Aaltonen et~al., {\it {Search for Higgs bosons
  predicted in two-Higgs-doublet models via decays to tau lepton pairs in 1.96
  TeV proton- antiproton collisions}},  {\em Phys. Rev. Lett.} {\bf 103} (2009)
  201801, [\href{http://arxiv.org/abs/0906.1014}{{\tt arXiv:0906.1014}}].

\bibitem{arXiv:1109.3357}
{\bf ATLAS} Collaboration, and others, {\it {Search for a Standard Model Higgs
  boson in the H->ZZ- llnunu decay channel with the ATLAS detector}},  {\em
  Phys. Rev. Lett.} {\bf 107} (2011) 221802,
  [\href{http://arxiv.org/abs/1109.3357}{{\tt arXiv:1109.3357}}].

\bibitem{arXiv:1202.1415}
{\bf ATLAS} Collaboration, G.~Aad et~al., {\it {Search for the Standard Model
  Higgs boson in the decay channel H->ZZ(*)->4l with 4.8 fb-1 of pp collision
  data at sqrt(s) = 7 TeV with ATLAS}},  {\em Phys. Lett.} {\bf B710} (2012)
  383--402, [\href{http://arxiv.org/abs/1202.1415}{{\tt arXiv:1202.1415}}].

\bibitem{hep-ex/0107032}
{\bf LEP Higgs Working for Higgs boson searches} Collaboration, {\it {Searches
  for Invisible Higgs bosons: Preliminary combined results using LEP data
  collected at energies up to 209 GeV}},
  \href{http://arxiv.org/abs/hep-ex/0107032}{{\tt hep-ex/0107032}}.

\bibitem{arXiv:1001.4468}
{\bf CDF} Collaboration, T.~Aaltonen et~al., {\it {Inclusive Search for
  Standard Model Higgs Boson Production in the WW Decay Channel using the CDF
  II Detector}},  {\em Phys. Rev. Lett.} {\bf 104} (2010) 061803,
  [\href{http://arxiv.org/abs/1001.4468}{{\tt arXiv:1001.4468}}].

\bibitem{arXiv:1109.3615}
{\bf ATLAS} Collaboration, and others, {\it {Search for the Higgs boson in the
  H->WW->lvjj decay channel in pp collisions at sqrt{s} = 7 TeV with the ATLAS
  detector}},  {\em Phys. Rev. Lett.} {\bf 107} (2011) 231801,
  [\href{http://arxiv.org/abs/1109.3615}{{\tt arXiv:1109.3615}}].

\bibitem{hep-ex/0401022}
{\bf DELPHI} Collaboration, J.~Abdallah et~al., {\it {Searches for invisibly
  decaying Higgs bosons with the DELPHI detector at LEP}},  {\em Eur. Phys. J.}
  {\bf C32} (2004) 475--492, [\href{http://arxiv.org/abs/hep-ex/0401022}{{\tt
  hep-ex/0401022}}].

\bibitem{arXiv:1409.6064}
{\bf ATLAS} Collaboration, {\it {Search for neutral Higgs bosons of the minimal
  supersymmetric standard model in pp collisions at $\sqrt{s}$ = 8 TeV with the
  ATLAS detector}},  \href{http://arxiv.org/abs/1409.6064}{{\tt
  arXiv:1409.6064}}.

\bibitem{hep-ex/0107031}
{\bf LEP Higgs Working Group for Higgs boson searches} Collaboration, {\it
  {Search for charged Higgs bosons: Preliminary combined results using LEP data
  collected at energies up to 209- GeV}},
  \href{http://arxiv.org/abs/hep-ex/0107031}{{\tt hep-ex/0107031}}.

\bibitem{arXiv:1207.0449}
{\bf Tevatron New Physics Higgs Working Group} Collaboration, C.~Group,
  D.~Collaborations, and the Tevatron New Physics~an, {\it {Updated Combination
  of CDF and D0 Searches for Standard Model Higgs Boson Production with up to
  10.0 fb$^{-1}$ of Data}},  \href{http://arxiv.org/abs/1207.0449}{{\tt
  arXiv:1207.0449}}.

\bibitem{arXiv:1507.05930}
{\bf ATLAS} Collaboration, {\it {Search for an additional, heavy Higgs boson in
  the $H\rightarrow ZZ$ decay channel at $\sqrt{s}$ = 8 TeV in $pp$ collision
  data with the ATLAS detector}},  \href{http://arxiv.org/abs/1507.05930}{{\tt
  arXiv:1507.05930}}.

\bibitem{arXiv:1106.4885}
{\bf D0} Collaboration, V.~M. Abazov et~al., {\it {Search for neutral Higgs
  bosons decaying to tau pairs produced in association with b quarks in ppbar
  collisions at sqrt(s)=1.96 TeV}},  {\em Phys. Rev. Lett.} {\bf 107} (2011)
  121801, [\href{http://arxiv.org/abs/1106.4885}{{\tt arXiv:1106.4885}}].

\bibitem{arXiv:0907.1269}
{\bf CDF} Collaboration, T.~Aaltonen et~al., {\it {Search for charged Higgs
  bosons in decays of top quarks in $p-\bar{p}$ collisions at $\sqrt{s} = 1.96$
  TeV}},  {\em Phys. Rev. Lett.} {\bf 103} (2009) 101803,
  [\href{http://arxiv.org/abs/0907.1269}{{\tt arXiv:0907.1269}}].

\bibitem{arXiv:1207.6436}
{\bf CDF} Collaboration, T.~Aaltonen et~al., {\it {Evidence for a particle
  produced in association with weak bosons and decaying to a bottom-antibottom
  quark pair in Higgs boson searches at the Tevatron}},  {\em Phys. Rev. Lett.}
  {\bf 109} (2012) 071804, [\href{http://arxiv.org/abs/1207.6436}{{\tt
  arXiv:1207.6436}}].

\bibitem{arXiv:1106.4555}
{\bf D0} Collaboration, V.~M. Abazov et~al., {\it {Search for Higgs bosons
  decaying to tautau pairs in ppbar collisions at sqrt(s) = 1.96 TeV}},  {\em
  Phys. Lett.} {\bf B707} (2012) 323--329,
  [\href{http://arxiv.org/abs/1106.4555}{{\tt arXiv:1106.4555}}].

\bibitem{arXiv:1003.3363}
{\bf Tevatron New Phenomena and Higgs Working Group} Collaboration, D.~Benjamin
  et~al., {\it {Combined CDF and D0 upper limits on MSSM Higgs boson production
  in tau-tau final states with up to 2.2 fb-1}},
  \href{http://arxiv.org/abs/1003.3363}{{\tt arXiv:1003.3363}}.

\bibitem{arXiv:0905.3381}
{\bf D0} Collaboration, V.~M. Abazov et~al., {\it {Search for NMSSM Higgs
  bosons in the $h \to a a \to\mu\mu\: \mu\mu, \mu\mu \: \tau\tau$ channels
  using $p \bar{p}$ collisions at $\sqrt{s}$=1.96 TeV}},  {\em Phys. Rev.
  Lett.} {\bf 103} (2009) 061801, [\href{http://arxiv.org/abs/0905.3381}{{\tt
  arXiv:0905.3381}}].

\bibitem{arXiv:1406.5053}
{\bf ATLAS} Collaboration, {\it {Search For Higgs Boson Pair Production in the
  $\gamma\gamma b\bar{b}$ Final State using $pp$ Collision Data at $\sqrt{s}=8$
  TeV from the ATLAS Detector}},  \href{http://arxiv.org/abs/1406.5053}{{\tt
  arXiv:1406.5053}}.

\bibitem{arXiv:1107.4960}
{\bf TEVNPH Working Group} Collaboration, and others, {\it {Combined CDF and D0
  Searches for the Standard Model Higgs Boson Decaying to Two Photons with up
  to 8.2 $fb^{-1}$}},  \href{http://arxiv.org/abs/1107.4960}{{\tt
  arXiv:1107.4960}}.

\bibitem{arXiv:0809.3930}
{\bf CDF} Collaboration, T.~Aaltonen et~al., {\it {Search for a Higgs Boson
  Decaying to Two W Bosons at CDF}},  {\em Phys. Rev. Lett.} {\bf 102} (2009)
  021802, [\href{http://arxiv.org/abs/0809.3930}{{\tt arXiv:0809.3930}}].

\bibitem{arXiv:1202.1414}
{\bf ATLAS} Collaboration, G.~Aad, {\it {Search for the Standard Model Higgs
  boson in the diphoton decay channel with 4.9 fb$^{-1}$ of $pp$ collisions at
  $\sqrt{s}=7$ TeV with ATLAS}},  {\em Phys. Rev. Lett.} {\bf 108} (2012)
  111803, [\href{http://arxiv.org/abs/1202.1414}{{\tt arXiv:1202.1414}}].

\bibitem{hep-ex/0602042}
{\bf ALEPH} Collaboration, S.~Schael et~al., {\it {Search for neutral MSSM
  Higgs bosons at LEP}},  {\em Eur. Phys. J.} {\bf C47} (2006) 547--587,
  [\href{http://arxiv.org/abs/hep-ex/0602042}{{\tt hep-ex/0602042}}].

\bibitem{arXiv:0906.5613}
{\bf CDF} Collaboration, T.~Aaltonen et~al., {\it {Search for a Higgs Boson in
  $W H \to \ell \nu b \bar{b}$ in $p\bar{p}$ Collisions at $\sqrt{s} = 1.96$
  TeV}},  {\em Phys. Rev. Lett.} {\bf 103} (2009) 101802,
  [\href{http://arxiv.org/abs/0906.5613}{{\tt arXiv:0906.5613}}].

\bibitem{arxiv:1202.1408}
{\bf ATLAS} Collaboration, G.~Aad, {\it {Combined search for the Standard Model
  Higgs boson using up to 4.9 fb$^{-1}$ of $pp$ collision data at $\sqrt{s}=7$
  TeV with the ATLAS detector at the LHC}},  {\em Phys. Lett.} {\bf B710}
  (2012) 49--66, [\href{http://arxiv.org/abs/1202.1408}{{\tt
  arXiv:1202.1408}}].

\bibitem{arXiv:1407.6583}
{\bf ATLAS} Collaboration, {\it {Search for Scalar Diphoton Resonances in the
  Mass Range $65-600$ GeV with the ATLAS Detector in $pp$ Collision Data at
  $\sqrt{s}$ = 8 $TeV$}},  \href{http://arxiv.org/abs/1407.6583}{{\tt
  arXiv:1407.6583}}.

\bibitem{arXiv:1509.00389}
{\bf ATLAS} Collaboration, {\it {Search for a high-mass Higgs boson decaying to
  a $W$ boson pair in $pp$ collisions at $\sqrt{s} = 8$ TeV with the ATLAS
  detector}},  \href{http://arxiv.org/abs/1509.00389}{{\tt arXiv:1509.00389}}.

\bibitem{hep-ex/0404012}
{\bf DELPHI} Collaboration, J.~Abdallah et~al., {\it {Search for charged Higgs
  bosons at LEP in general two Higgs doublet models}},  {\em Eur. Phys. J.}
  {\bf C34} (2004) 399--418, [\href{http://arxiv.org/abs/hep-ex/0404012}{{\tt
  hep-ex/0404012}}].

\bibitem{arXiv:0901.1887}
{\bf D0} Collaboration, V.~M. Abazov et~al., {\it {Search for Resonant Diphoton
  Production with the D0 Detector}},  {\em Phys. Rev. Lett.} {\bf 102} (2009)
  231801, [\href{http://arxiv.org/abs/0901.1887}{{\tt arXiv:0901.1887}}].

\bibitem{CDFnotes}
{\bf CDF} Collaboration.
\newblock CDF Notes 10500 7307 10439 10796 9999 10485 10798 8353 10799 10599
  10573 10010 7712 10574.

\bibitem{D0notes}
{\bf D0} Collaboration.
\newblock D0 Notes 6083 6305 6227 6299 6301 6302 5739 5845 6286 5757 6296 6183
  6295 6171 6309 6276 6304 5873.

\bibitem{CMSnotes}
{\bf CMS} Collaboration.
\newblock CMS Physics Analysis Summaries.

\bibitem{ATLASnotes}
{\bf ATLAS} Collaboration.
\newblock ATLAS CONF Notes 2012-160 2014-049 2012-135 2012-161 2012-092
  2012-018 2012-012 2013-010 2013-013 2012-019 2012-168 2012-078 2014-050
  2012-017 2012-016 2011-094 2013-030 2011-157.

\bibitem{LHWGnotes}
{\bf LHWG} Collaboration.
\newblock LHWG Notes 2002-02.

\bibitem{hep-ph/9704448}
A.~Djouadi, J.~Kalinowski, and M.~Spira, {\it {HDECAY: A program for Higgs
  boson decays in the standard model and its supersymmetric extension}},  {\em
  Comput. Phys. Commun.} {\bf 108} (1998) 56--74,
  [\href{http://arxiv.org/abs/hep-ph/9704448}{{\tt hep-ph/9704448}}].

\bibitem{hep-ph/0102227}
S.~Catani, D.~de~Florian, and M.~Grazzini, {\it {Higgs production in hadron
  collisions: Soft and virtual QCD corrections at NNLO}},  {\em JHEP} {\bf 05}
  (2001) 025, [\href{http://arxiv.org/abs/hep-ph/0102227}{{\tt
  hep-ph/0102227}}].

\bibitem{hep-ph/0102241}
R.~V. Harlander and W.~B. Kilgore, {\it {Soft and virtual corrections to p p
  --> H + X at NNLO}},  {\em Phys. Rev.} {\bf D64} (2001) 013015,
  [\href{http://arxiv.org/abs/hep-ph/0102241}{{\tt hep-ph/0102241}}].

\bibitem{hep-ph/0201206}
R.~V. Harlander and W.~B. Kilgore, {\it {Next-to-next-to-leading order Higgs
  production at hadron colliders}},  {\em Phys. Rev. Lett.} {\bf 88} (2002)
  201801, [\href{http://arxiv.org/abs/hep-ph/0201206}{{\tt hep-ph/0201206}}].

\bibitem{hep-ph/0207004}
C.~Anastasiou and K.~Melnikov, {\it {Higgs boson production at hadron colliders
  in NNLO QCD}},  {\em Nucl. Phys.} {\bf B646} (2002) 220--256,
  [\href{http://arxiv.org/abs/hep-ph/0207004}{{\tt hep-ph/0207004}}].

\bibitem{hep-ph/0302135}
V.~Ravindran, J.~Smith, and W.~L. van Neerven, {\it {NNLO corrections to the
  total cross section for Higgs boson production in hadron hadron collisions}},
   {\em Nucl. Phys.} {\bf B665} (2003) 325--366,
  [\href{http://arxiv.org/abs/hep-ph/0302135}{{\tt hep-ph/0302135}}].

\bibitem{arXiv:0811.3458}
C.~Anastasiou, R.~Boughezal, and F.~Petriello, {\it {Mixed QCD-electroweak
  corrections to Higgs boson production in gluon fusion}},  {\em JHEP} {\bf 04}
  (2009) 003, [\href{http://arxiv.org/abs/0811.3458}{{\tt arXiv:0811.3458}}].

\bibitem{Dawson:1990zj}
S.~Dawson, {\it {Radiative corrections to Higgs boson production}},  {\em Nucl.
  Phys.} {\bf B359} (1991) 283--300.

\bibitem{Djouadi:1991tka}
A.~Djouadi, M.~Spira, and P.~M. Zerwas, {\it {Production of Higgs bosons in
  proton colliders: QCD corrections}},  {\em Phys. Lett.} {\bf B264} (1991)
  440--446.

\bibitem{hep-ph/9504378}
M.~Spira, A.~Djouadi, D.~Graudenz, and P.~M. Zerwas, {\it {Higgs boson
  production at the LHC}},  {\em Nucl. Phys.} {\bf B453} (1995) 17--82,
  [\href{http://arxiv.org/abs/hep-ph/9504378}{{\tt hep-ph/9504378}}].

\bibitem{hep-ph/0404071}
U.~Aglietti, R.~Bonciani, G.~Degrassi, and A.~Vicini, {\it {Two-loop light
  fermion contribution to Higgs production and decays}},  {\em Phys. Lett.}
  {\bf B595} (2004) 432--441, [\href{http://arxiv.org/abs/hep-ph/0404071}{{\tt
  hep-ph/0404071}}].

\bibitem{hep-ph/0407249}
G.~Degrassi and F.~Maltoni, {\it {Two-loop electroweak corrections to Higgs
  production at hadron colliders}},  {\em Phys. Lett.} {\bf B600} (2004)
  255--260, [\href{http://arxiv.org/abs/hep-ph/0407249}{{\tt hep-ph/0407249}}].

\bibitem{arXiv:0809.1301}
S.~Actis, G.~Passarino, C.~Sturm, and S.~Uccirati, {\it {NLO Electroweak
  Corrections to Higgs Boson Production at Hadron Colliders}},  {\em Phys.
  Lett.} {\bf B670} (2008) 12--17, [\href{http://arxiv.org/abs/0809.1301}{{\tt
  arXiv:0809.1301}}].

\bibitem{arXiv:0809.3667}
S.~Actis, G.~Passarino, C.~Sturm, and S.~Uccirati, {\it {NNLO Computational
  Techniques: the Cases $H \to \gamma \gamma$ and $H \to g g$}},  {\em Nucl.
  Phys.} {\bf B811} (2009) 182--273,
  [\href{http://arxiv.org/abs/0809.3667}{{\tt arXiv:0809.3667}}].

\bibitem{hep-ph/0306211}
S.~Catani, D.~de~Florian, M.~Grazzini, and P.~Nason, {\it {Soft-gluon
  resummation for Higgs boson production at hadron colliders}},  {\em JHEP}
  {\bf 07} (2003) 028, [\href{http://arxiv.org/abs/hep-ph/0306211}{{\tt
  hep-ph/0306211}}].

\bibitem{arXiv:0901.2427}
D.~de~Florian and M.~Grazzini, {\it {Higgs production through gluon fusion:
  updated cross sections at the Tevatron and the LHC}},  {\em Phys. Lett.} {\bf
  B674} (2009) 291--294, [\href{http://arxiv.org/abs/0901.2427}{{\tt
  arXiv:0901.2427}}].

\bibitem{hep-ph/0307206}
O.~Brein, A.~Djouadi, and R.~Harlander, {\it {NNLO QCD corrections to the
  Higgs-strahlung processes at hadron colliders}},  {\em Phys. Lett.} {\bf
  B579} (2004) 149--156, [\href{http://arxiv.org/abs/hep-ph/0307206}{{\tt
  hep-ph/0307206}}].

\bibitem{hep-ph/0306234}
M.~L. Ciccolini, S.~Dittmaier, and M.~Kramer, {\it {Electroweak radiative
  corrections to associated W H and Z H production at hadron colliders}},  {\em
  Phys. Rev.} {\bf D68} (2003) 073003,
  [\href{http://arxiv.org/abs/hep-ph/0306234}{{\tt hep-ph/0306234}}].

\bibitem{hep-ph/0406152}
{\bf Higgs Working Group} Collaboration, K.~A. Assamagan et~al., {\it {The
  Higgs working group: Summary report 2003}},
  \href{http://arxiv.org/abs/hep-ph/0406152}{{\tt hep-ph/0406152}}.

\bibitem{hep-ph/0304035}
R.~V. Harlander and W.~B. Kilgore, {\it {Higgs boson production in bottom quark
  fusion at next-to- next-to-leading order}},  {\em Phys. Rev.} {\bf D68}
  (2003) 013001, [\href{http://arxiv.org/abs/hep-ph/0304035}{{\tt
  hep-ph/0304035}}].

\bibitem{hep-ph/9206246}
T.~Han, G.~Valencia, and S.~Willenbrock, {\it {Structure function approach to
  vector boson scattering in p p collisions}},  {\em Phys. Rev. Lett.} {\bf 69}
  (1992) 3274--3277, [\href{http://arxiv.org/abs/hep-ph/9206246}{{\tt
  hep-ph/9206246}}].

\bibitem{hep-ph/9905386}
J.~M. Campbell and R.~K. Ellis, {\it {An update on vector boson pair production
  at hadron colliders}},  {\em Phys. Rev.} {\bf D60} (1999) 113006,
  [\href{http://arxiv.org/abs/hep-ph/9905386}{{\tt hep-ph/9905386}}].

\bibitem{hep-ph/0306109}
T.~Figy, C.~Oleari, and D.~Zeppenfeld, {\it {Next-to-leading order jet
  distributions for Higgs boson production via weak-boson fusion}},  {\em Phys.
  Rev.} {\bf D68} (2003) 073005,
  [\href{http://arxiv.org/abs/hep-ph/0306109}{{\tt hep-ph/0306109}}].

\bibitem{hep-ph/0403194}
E.~L. Berger and J.~M. Campbell, {\it {Higgs boson production in weak boson
  fusion at next-to- leading order}},  {\em Phys. Rev.} {\bf D70} (2004)
  073011, [\href{http://arxiv.org/abs/hep-ph/0403194}{{\tt hep-ph/0403194}}].

\bibitem{hep-ph/0612172}
U.~Aglietti et~al., {\it {Tevatron for LHC report: Higgs}},
  \href{http://arxiv.org/abs/hep-ph/0612172}{{\tt hep-ph/0612172}}.

\bibitem{hep-ph/0107081}
W.~Beenakker et~al., {\it {Higgs radiation off top quarks at the Tevatron and
  the LHC}},  {\em Phys. Rev. Lett.} {\bf 87} (2001) 201805,
  [\href{http://arxiv.org/abs/hep-ph/0107081}{{\tt hep-ph/0107081}}].

\bibitem{hep-ph/0107101}
L.~Reina and S.~Dawson, {\it {Next-to-leading order results for t anti-t h
  production at the Tevatron}},  {\em Phys. Rev. Lett.} {\bf 87} (2001) 201804,
  [\href{http://arxiv.org/abs/hep-ph/0107101}{{\tt hep-ph/0107101}}].

\bibitem{hep-ph/0211438}
S.~Dawson, L.~H. Orr, L.~Reina, and D.~Wackeroth, {\it {Associated top quark
  Higgs boson production at the LHC}},  {\em Phys. Rev.} {\bf D67} (2003)
  071503, [\href{http://arxiv.org/abs/hep-ph/0211438}{{\tt hep-ph/0211438}}].

\bibitem{hep-ph/0305321}
O.~Brein and W.~Hollik, {\it {MSSM Higgs bosons associated with high-p(T) jets
  at hadron colliders}},  {\em Phys. Rev.} {\bf D68} (2003) 095006,
  [\href{http://arxiv.org/abs/hep-ph/0305321}{{\tt hep-ph/0305321}}].

\bibitem{arXiv:0705.2744}
O.~Brein and W.~Hollik, {\it {Distributions for MSSM Higgs boson + jet
  production at hadron colliders}},  {\em Phys. Rev.} {\bf D76} (2007) 035002,
  [\href{http://arxiv.org/abs/0705.2744}{{\tt arXiv:0705.2744}}].

\bibitem{arXiv:0707.0381}
M.~Ciccolini, A.~Denner, and S.~Dittmaier, {\it {Strong and electroweak
  corrections to the production of Higgs+2jets via weak interactions at the
  LHC}},  {\em Phys. Rev. Lett.} {\bf 99} (2007) 161803,
  [\href{http://arxiv.org/abs/0707.0381}{{\tt arXiv:0707.0381}}].

\bibitem{arXiv:0710.4749}
M.~Ciccolini, A.~Denner, and S.~Dittmaier, {\it {Electroweak and QCD
  corrections to Higgs production via vector-boson fusion at the LHC}},  {\em
  Phys. Rev.} {\bf D77} (2008) 013002,
  [\href{http://arxiv.org/abs/0710.4749}{{\tt arXiv:0710.4749}}].

\bibitem{arXiv:1101.0593}
{\bf LHC Higgs Cross Section Working Group} Collaboration, S.~Dittmaier et~al.,
  {\it {Handbook of LHC Higgs Cross Sections: 1. Inclusive Observables}},
  \href{http://arxiv.org/abs/1101.0593}{{\tt arXiv:1101.0593}}.

\bibitem{arXiv:1201.3084}
S.~Dittmaier et~al., {\it {Handbook of LHC Higgs Cross Sections: 2.
  Differential Distributions}},  \href{http://arxiv.org/abs/1201.3084}{{\tt
  arXiv:1201.3084}}.

\bibitem{arXiv:1307.1347}
T.~L. H. C. S.~W. Group et~al., {\it {Handbook of LHC Higgs Cross Sections: 3.
  Higgs Properties}},  \href{http://arxiv.org/abs/1307.1347}{{\tt
  arXiv:1307.1347}}.

\bibitem{Buckley:2013jua}
A.~Buckley, {\it {PySLHA: a Pythonic interface to SUSY Les Houches Accord
  data}},  {\em Eur. Phys. J.} {\bf C75} (2015), no.~10 467,
  [\href{http://arxiv.org/abs/1305.4194}{{\tt arXiv:1305.4194}}].

\bibitem{Harlander:2012pb}
R.~V. Harlander, S.~Liebler, and H.~Mantler, {\it {SusHi: A program for the
  calculation of Higgs production in gluon fusion and bottom-quark annihilation
  in the Standard Model and the MSSM}},  {\em Comput. Phys. Commun.} {\bf 184}
  (2013) 1605--1617, [\href{http://arxiv.org/abs/1212.3249}{{\tt
  arXiv:1212.3249}}].

\bibitem{Harlander:2016hcx}
R.~V. Harlander, S.~Liebler, and H.~Mantler, {\it {SusHi Bento: Beyond NNLO and
  the heavy-top limit}},  {\em Comput. Phys. Commun.} {\bf 212} (2017)
  239--257, [\href{http://arxiv.org/abs/1605.03190}{{\tt arXiv:1605.03190}}].

\bibitem{Haisch:2016usn}
U.~Haisch, F.~Kahlhoefer, and T.~M.~P. Tait, {\it {On Mono-W Signatures in
  Spin-1 Simplified Models}},  {\em Phys. Lett.} {\bf B760} (2016) 207--213,
  [\href{http://arxiv.org/abs/1603.01267}{{\tt arXiv:1603.01267}}].

\bibitem{Barr:2016inz}
A.~Barr and J.~Liu, {\it {First interpretation of 13 TeV supersymmetry searches
  in the pMSSM}},  \href{http://arxiv.org/abs/1605.09502}{{\tt
  arXiv:1605.09502}}.

\bibitem{Barr:2016sho}
A.~Barr and J.~Liu, {\it {Analysing parameter space correlations of recent 13
  TeV gluino and squark searches in the pMSSM}},  {\em Eur. Phys. J.} {\bf C77}
  (2017), no.~3 202, [\href{http://arxiv.org/abs/1608.05379}{{\tt
  arXiv:1608.05379}}].

\bibitem{pmssm:barr}
{http://www-pnp.physics.ox.ac.uk/~jesseliu/pmssm/}.

\end{thebibliography}\endgroup
\end{document}